\documentclass[useAMS,usenatbib]{mnras}
\usepackage{graphicx,natbib,amssymb,amsmath,stmaryrd,makecell}
\usepackage{txfonts}
\usepackage{xcolor}
\usepackage{hyperref}
\usepackage{xspace}

\newcommand{\expf}[1]{{{\rm e}^{#1}}}

\newcommand{\id}{{\,\rm d}}

\newcommand{\beq}{\begin{equation}}   %

\newcommand{\eeq}{\end{equation}}   %

\newcommand{\beqa}{\begin{eqnarray}}   %

\newcommand{\eeqa}{\end{eqnarray}}   %

\newcommand{\beal}{\begin{align}}

\newcommand{\enal}{\end{align}}

\newcommand{\bspl}{\begin{split}}

\newcommand{\espl}{\end{split}}

\newcommand{\bsub}{\begin{subequations}}

\newcommand{\esub}{\end{subequations}}

\newcommand{\bmulti}{\begin{multline}}   %

\newcommand{\beqm}{\begin{mathletters}}   %

\newcommand{\eeqm}{\end{mathletters}}   %

\newcommand{\pot}[2]{#1 \times 10^{#2}}

\newcommand{\COBEF}{{\it COBE/FIRAS}\xspace}

\title[radio excess from dark photons]{A closer look at dark photon explanations of the excess radio background}

\author[Acharya and Chluba]
{Sandeep Kumar Acharya$^1$\thanks{E-mail:sandeep.acharya@manchester.ac.uk} 
and
Jens Chluba$^1$
\\
$^1$Jodrell Bank Centre for Astrophysics, School of Physics and Astronomy, The University of Manchester, Manchester M13 9PL, U.K.
}
\date{\vspace{-5mm}Accepted XXX. Received YYY; in original form ZZZ}

\begin{document}

\maketitle

\begin{abstract}
The observed excess radio background has remained a puzzle for over a decade. A recent new physics solution involves dark matter that decays into dark photons in the presence of a thermal dark photon background. The produced non-thermal dark photon spectrum then converts into standard photons around the reionization era, yielding an approximate power-law radio excess with brightness temperature $T(\nu)\simeq \nu^{-2.5}$ over a wide range of frequencies, $\nu$. This simple power-law model comes intriguingly close to the current data, even if several ingredients are required to make it work. In this paper, we investigate some of the details of this model, showcasing the importance of individual effects. In particular, significant deviation from a power law are present at $\nu\lesssim 100\,{\rm MHz}$ and $\nu\gtrsim 1\,{\rm GHz}$. These effects result in improving the fit to data compared to a power-law spectrum, and may become testable in future observations. We also highlight independent signatures that can be tested with future CMB spectral distortion experiments such as {\it PIXIE}. 
However, there are challenges for the model from the observed radio background anisotropies, as discussed here. 
We furthermore highlight a possible runaway process due to the finite width of the dark matter decay profile, which suggests that additional work might be required to obtain a viable model.
\end{abstract}

\begin{keywords}
cosmology - cosmic background radiation; cosmology - dark matter; cosmology - theory
\vspace{-5mm}
\end{keywords}

\section{Introduction}
The detection of an excess radio monopole at $\simeq 0.1-10$~GHz \citep{ARCADE2011,DT2018} on top of the Cosmic Microwave Background (CMB) is one of the intriguing astrophysical observations of the recent times. Although one expects a radio background from unresolved extra-galactic radio sources \citep{PB1996}, it is at least a factor of $\simeq 3$ smaller than the detected excess \citep{GTZBS2008}. In addition, the detection of 21cm absorption feature around $z\simeq 20$ \citep{Edges2018} has further added to this puzzle \citep{FH2018}, though the signal is yet to be confirmed \citep{Saras2022}. It was shown in \cite{ARCADE2011} that the radio synchrotron background (RSB) data (which includes ARCADE data) is well-fit by a power-law with slope $\gamma\simeq -2.6$ and temperature $T \simeq 24.1$~K at 310 MHz. The authors in \cite{DT2018} used independent data together with the re-analyzed ARCADE data points to find similar results with a consistent slope but slightly higher normalization of $T\simeq 30$~K at 310 MHz. 

A possible, conservative explanation for the radio excess would be a population of faint, undetected radio sources which are hosted by dark matter halos. Since dark matter halos are clustered and have an anisotropic distribution \citep{S2000,SPJWFPTEC2003}, one naturally expects the radio background to also be anisotropic. However, this possibility was shown to run into significant constraints \citep{Holder2014} even if recent results show that the level of radio background anisotropy is a complicated question to answer \citep{OSHHL2022}. Finding an explanation for both the excess radio monopole and absence of anisotropy has been a major theoretical challenge, and there still are open questions regarding the analysis procedure for the subtraction of the galactic contribution \citep{SC2013}.
For a detailed discussion on this topic, the readers are referred to \cite{Singal2018}.

New physics explanations for the radio excess of cosmological origin could be related to Comptonized photon injection distortions \citep{Chluba2015GreensII, Bolliet2020PI}, annihilating axion-like dark matter \citep{Fraser2018}, dark photon conversion \citep{PPRU2018, CLMPR2022}, superconducting cosmic strings \citep{BCS2019}, decay of relic neutrinos to sterile neutrinos \citep{CDFS2018} or the thermal emission of quark nugget dark matter \citep{LZ2019}. Alternative astrophysical solutions consider supernova explosions of population III stars \citep{JNB2019}, bright luminous galaxies \citep{MF2019} and accreting astrophysical \citep{ECLDSM2018,ECL2020} or primordial \citep{MK2021} black holes. 

Astrophysical sources that provide radio photons, in general, also emit UV/X-ray photons. These energetic photons can modify the thermal history of the Universe and, therefore, are constrained by various cosmological probes \citep{ADC2022}. There also are interesting ways to detect the presence of a radio background which uses the up-scattering of CMB photons by hot electrons inside galaxy clusters to predict the distortion to CMB blackbody spectrum in the radio bands \citep{HC2021,Lee2022rSZ}. Together with X-ray observations, these could provide a litmus test for the origin of the radio excess.

Recently, a dark sector model was proposed \citep[][henceforth C22]{CLMPR2022} to explain both the amplitude of radio monopole as well as its spectral and spatial smoothness. The model has three key ingredients: (1) dark matter decaying to non-thermal dark photons in a matter-dominated universe, (2) presence of a thermal dark photon background which stimulates the decay of dark matter to non-thermal dark photons and (3) resonant conversion of dark photons to standard model photons predominantly at redshift $z\lesssim$ 6-7. 
All these ingredients could be individually disputed; 
however, together they lead to the prediction of radio excess with an approximate power-law $T\propto \nu^{-5/2}$. Even if the power-law index of $\gamma=-5/2$ is a bit lower than the best-fit slope to the radio excess data, it is intriguingly close, and C22 argued that it provides a reasonable fit. While there are other cosmological constraints \citep{MRS2009,BPS2020,WRMP2020,AABPPS2020,CLMR20201} on model parameters of dark photon to photon conversion, it was shown that a reasonable range of parameters able to explain the radio excess seems to exist.  

In this paper, we study individual aspects of the model proposed by C22. Our approach is to  showcase the importance of these details which can give rise to  signatures that can be tested in the future and also improves the understanding of the cumulative results presented in C22. As an example, the contribution of radiation and dark energy to the Hubble parameter, even in a matter dominated universe, adds curvature to the radio excess signal at $\nu\lesssim$~100 MHz and $\nu\gtrsim 1$~GHz. Additional aspects such as using the exact stimulated decay factor and frequency-dependent conversion probability of dark photons to photons make the signal even more curved. We highlight that the curved spectrum is a better fit to the ARCADE data from the dark photon induced radio background alone when compared to the power-law approximation. Including a minimal extra-galactic radio background (MEG), we confirm that both the curved and power-law spectrum provide reasonable fit to the data. We argue that more precise data at $\simeq 1-10$ GHz will be able to test the validity of this model. We further discuss additional predictions of this model in terms of spectral distortion within the CMB bands (60-600 GHz), which could be tested using proposed spectrometers such as {\it PIXIE} \citep{Kogut2011PIXIE,Kogut2016SPIE}. 

However, there may still be challenges to this model in terms of explaining the spatial smoothness of this radio background which need to be overcome.  
In Sect.~\ref{sec:stim}, we also briefly discuss the possibility of self-stimulated decay, mediated by the created non-thermal photon population, which we find to change the soft photon spectrum significantly, indicating that additional work may be required to obtain a consistent model.

The paper is structured as follows: In Sec.~\ref{sec:overview}, we give a brief overview of the key arguments and calculations of C22. We detail several aspects of this model in Sec.~\ref{sec:main}. Our main results can be found in Sec \ref{sec:reanalysis}. In Sect.~\ref{sec:spectral_distortion}, we predict the possible spectral distortion signal within the CMB bands, which may be detected in future. We discuss possible challenges to the model in Sec.~\ref{sec:limitations} and close with a discussion in Sect.~\ref{sec:discussion}.

\vspace{-5mm}
\section{Brief overview of calculation of C22}
\label{sec:overview}
\begin{figure}
\centering 
\includegraphics[width=0.95\columnwidth]{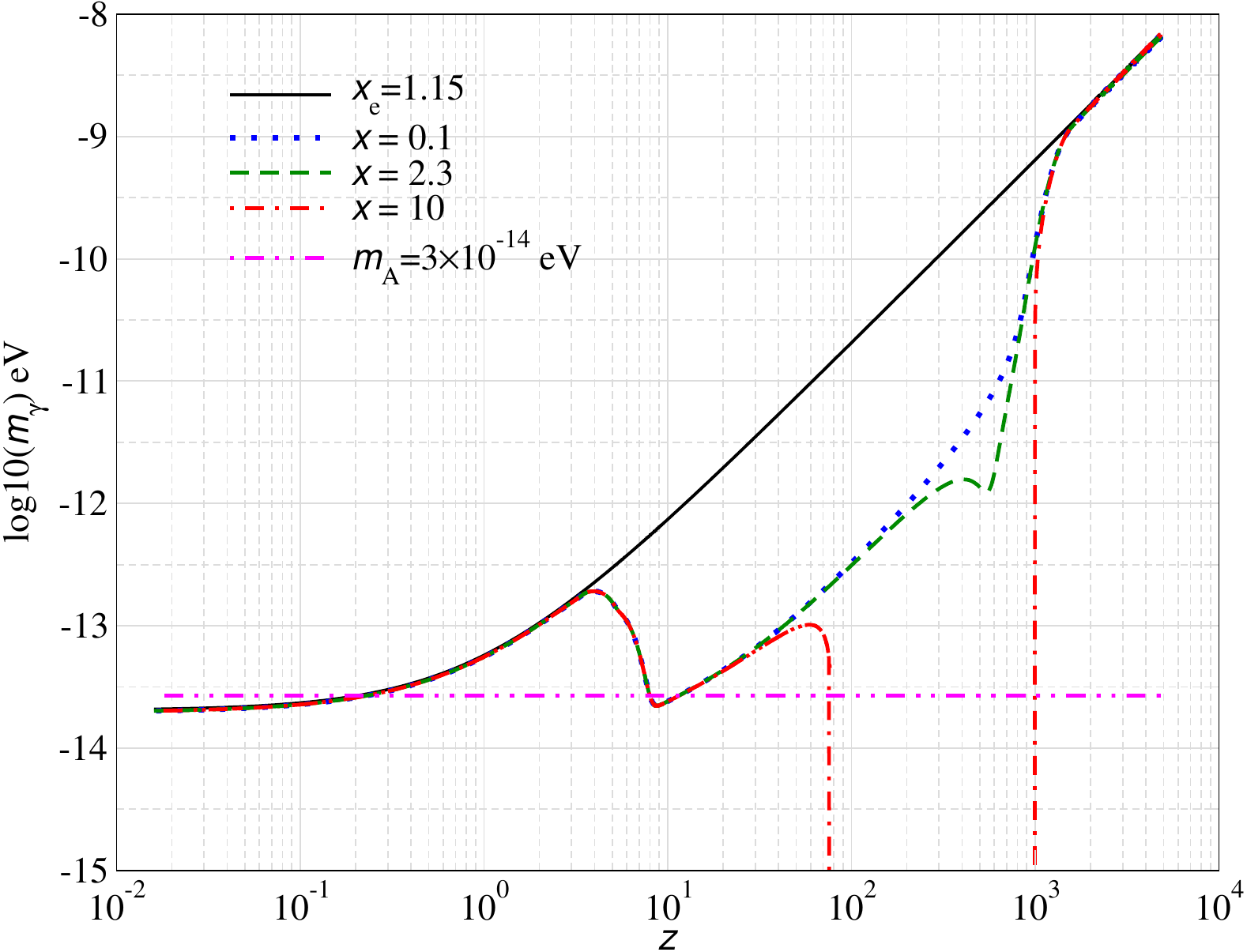}
\caption{Effective photon mass as a function of redshift for different frequencies with electron number density given by {\tt CosmoRec} \citep{Chluba2010b}. For $x=10$, the photon mass becomes imaginary at $10^2\lesssim z\lesssim 10^3$ and Eq.~\eqref{eq:m_gamma} breaks down. However, We will be interested in $x\lesssim 0.1$ and $z\lesssim 6$ when the Universe is reionized. We also indicate the mass of dark photon at $m_{A}=3\times10^{-14}$ eV, which is the median of the allowed $m_A$ parameter space obtained in C22. At frequency $x \ll 1$, the effective photon mass becomes independent of the photon frequency. For reference, we also show a hypothetical case of a completely ionized universe.}
\label{fig:mA}
\vspace{-3mm}
\end{figure}

In this section, we briefly describe the calculation of C22. We consider a dark matter particle 'X' with mass $m_X$. This particle decays to two particles, one or both of which can be a dark photon. The energy of the dark photon is $\epsilon_A=\frac{m_X}{2}$. The dark photon resonantly converts into a standard model photon when the dark photon mass, $m_A$, fulfills the resonance condition, $m_{A}\approx \bar{m}^*_{\gamma}$, where $\bar{m}^*_{\gamma}$ is the effective mass of standard model photon in the homogeneous universe. The effective photon mass is given by \citep[e.g.,][]{MRS2009}
\begin{align}
    \bar{m}^{*2}_{\gamma}(t)
    &\approx 1.4\times 10^{-21} {\rm eV}^2\,\frac{n_{\rm e}(t)}{\rm cm^{-3}}-8.4\times 10^{-24} {\rm eV}^2\left[\frac{\nu(t)}{\rm eV}\right]^2 \frac{n_{\rm HI}(t)}{\rm cm^{-3}}
    \nonumber\\
    &\approx 
    1.4\times 10^{-21} {\rm eV}^2\,\frac{n_{\rm e}(t)}{\rm cm^{-3}}
    \left[1
    \!-\!3.3\times 10^{-10}\, \frac{x^2}{a^2} \frac{n_{\rm HI}(t)}{n_{\rm e}(t)}
    \right],
    \label{eq:m_gamma}
\end{align}
where $n_{\rm e}$ is the free electron number density, $n_{\rm HI}$ is the neutral hydrogen number density, $\nu$ is the energy of the standard photon and $x=\nu/T_{\rm CMB}(z)$ is the dimensionless and redshift-independent photon frequency in terms of the CMB temperature, $T_{\rm CMB}=T_0(1+z)$ at redshift $z$, with present-day temperature $T_0$. We also used $a=1/(1+z)$ for the scale factor. 
We are interested in scenarios for which dark photons convert into photons at $z\lesssim 6.5$, as we explain below. This means that $m_{A}$ has to be in the mass range $\simeq 10^{-14}-5\times 10^{-13}$~eV to satisfy the criterion (see Fig.~\ref{fig:mA}). 

In order to explain the power-law fit to the radio excess data \citep{ARCADE2011,DT2018}, in the model proposed by C22 it is required that the dark photons converting into photons within the frequency range 22 MHz$-$10 GHz (as seen today) are produced in the matter-dominated era. This tightly constrains the allowed parameters space of $m_X$. 

To understand this range, we need to distinguish three redshifts: $z_*$ shall denote the redshift at which the dark photon is produced by the dark matter decay; $z'$ shall be the redshift at which the dark photon converts into the standard model photon; and $z$ is the redshift at which we observe the standard model photon. At the time of conversion, the standard model photon frequency is determined by 
\begin{equation}
\label{eq:nu_nuA}
    \nu'=\nu_{A,0} \left(\frac{1+z'}{1+z_*}\right),
\end{equation}
where $\nu_{A,0}\approx \epsilon_A$ (for $m_A\ll \epsilon_A=m_X/2$) is the dark photon frequency. The frequency of the standard photon observed at a later redshift $z$ is then $\nu=\nu' (1+z)/(1+z')=\nu_{A,0}(1+z)/(1+z_*)$, as expected.

The requirement for the dark photon to be produced in the matter-dominated era implies $z_\Lambda \lesssim z_*\lesssim z_{\rm eq}$, where $z_\Lambda\approx 0.3$ is the redshift at which the cosmological constant starts dominating and $z_{\rm eq}\simeq 3400$ is the redshift of matter-radiation equality \citep{Planck2018params}.\footnote{To steer clear of the transition regimes, one should limit $z_*$ further, an aspect that leads to corrections as we demonstrate below.} 
On the other hand, the range of standard photon frequencies we aim to span is $22~{\rm MHz}\lesssim \nu\lesssim 10\,{\rm GHz}$. Assuming that the conversion occurs at one single redshift, this constrains the upper and lower boundary for the dark photon production redshift to $(1+z_*^{\rm up})/(1+z_*^{\rm low})\simeq 10\,{\rm GHz}/22~{\rm MHz}\simeq 455$. Assuming the extreme case that $z_*^{\rm low}=z'=0$, we then have $z_*^{\rm up}\approx 454$ and $\nu_{A,0}\approx 10\,{\rm GHz}$ or $m_X=\pot{8.3}{-5}\,{\rm eV}$. On the other hand, for dark photon production at $z_*^{\rm up}=3400$, we have $z_*^{\rm low}\approx 6.5$ and $\nu_{A,0}\approx 10\,{\rm GHz}\,(1+6.5)\approx 75\,\,{\rm GHz}$ or $m_X=\pot{6.2}{-4}\,{\rm eV}$. 
Put together, this means $8\times 10^{-5}\lesssim m_X \lesssim 6\times 10^{-4}$~eV, which also implies $m_{A}\ll m_X$, guaranteeing that the dark photons are relativistic. 
In addition, the conversion redshift has to fulfill $z'\lesssim 6.5$.
The requirement that the standard photons, spanning 3 decades in frequency, must be emitted in the matter-dominated era, therefore sets stringent constraints on the dark matter-dark photon model parameters. 

The C22 model also requires a blackbody distribution of the same dark photon candidate, which is assumed to be undistorted throughout the evolution. This leads to enhancement of decay lifetime and is a critical ingredient for explaining the power-law index of $\gamma=-2.5$. The rate of stimulated decay rate is given by,
\begin{equation}
    \Gamma^*_X(z)=(1.0+nf_{A})\Gamma_X,
\end{equation}
with dark photon blackbody occupation, $f_{A}=\frac{1}{e^{\nu_A/T_A}-1}$ and $n=2$ for bosonic particles \citep[e.g.,][for explanation]{Bolliet2020PI}. In vacuum, the lifetime is given by the inverse decay rate, $t_X=1/\Gamma_X$. The temperature of background dark photon distribution is given by $T_A=T_{A,0}(1+z)$ and $T_{A,0}$ is the temperature of the thermal dark photon background as seen today. For $\nu_A \ll T_A$, the stimulated decay rate is given by $\Gamma^*_X(z)\approx\frac{2 T_{A}}{\nu_A}\Gamma_X$. At redshift $z$, the dark photon number density per frequency is then given by (see Appendix~\ref{app:emission_solution}),
\begin{equation}
\label{eq:sol_C22}
    \frac{{\rm d}\tilde{N}_{A}(z)}{{\rm d}\nu_A}\approx\frac{{\rm c}\alpha}{4\pi\nu_A}\frac{\tilde{N}_X\Gamma_X^*(z_*)}{H(z_*)}\Theta(z_*-z),
\end{equation}
where $\tilde{N}_A$ and $\tilde{N}_X$ are comoving number density of dark photons and dark matter respectively and $\alpha=1$ or 2 if one or both decay products are a dark photon and $\tilde{N}_X$ denotes comoving dark matter number density.

Once the resonance condition for dark photon to standard model photon conversion is met, we will have a photon distribution with same frequency dependence as that of the dark photons. The coupling constant for this conversion is given by $\epsilon$ [details can be found in Eq.~\eqref{eq:dP_dz}] and is of the order of $10^{-6}-10^{-8}$ in order to explain the radio excess (C22). 
The number density of standard photon per frequency is then given by (C22),
\begin{equation}
    \frac{{\rm d}\tilde{N}_{\gamma}(z,\nu)}{{\rm d}\nu}= \int^{z_*}_z {\rm d}z' \frac{{\rm d}P_{A\rightarrow \gamma}(z')}{{\rm d}z'} \frac{{\rm d}\tilde{N}_{A}(z')}{{\rm d}\nu_A},
\end{equation}
where $\int^{z_*}_z {\rm d}z' \frac{{\rm d}P_{A\rightarrow \gamma}(z')}{{\rm d}z'}$ is the probability of conversion of $A$ into standard photons, which scales $\propto 1/x$. We also used Eq.~\eqref{eq:nu_nuA} to convert between $\nu_{A,0}$ and $\nu$.
For a matter-dominated universe, one has $H(z_*)\propto (1+z_*)^{3/2}\propto x^{-3/2}$ and therefore $\frac{{\rm d}N_{\gamma}}{{\rm d}x}\propto x^{3/2}/x^3= x^{-3/2}$. This implies a radio brightness temperature of $T \propto (x \frac{{\rm d}N_{\gamma}}{{\rm d}x})/x^2 \propto x^{-5/2}$.

Putting everything together, the brightness temperature of the radio excess is approximately given by (C22)\footnote{The expression given in C22 also states dependence on $\nu_{A,0}$; however, in the model $\nu_{A,0}$ is not an independent parameter and fixed by $\nu_{A,0}=m_A/2$.},
\begin{align}
    T&\approx 10\,{\rm K} \left(\frac{\nu}{\nu_0}\right)^{-2.5} \left[\frac{T_{A,0}/T_0}{0.2}\right]\left[\frac{10^{21}{\rm s}}{\tau_X}\right]
    \left[\frac{2\times 10^{-4}{\rm eV}}{m_X}\right]^{5/2} \left[\frac{P_{\rm GHz}}{10^{-5}}\right],
     \label{eq:temp_amplitude}
\end{align}
where $\nu_0=310 \,{\rm MHz}$ and $P_{\rm GHz}=\int^{\infty}_0 {\rm d}z' \frac{{\rm d}P_{A\rightarrow \gamma}(z')}{{\rm d}z'}$ denotes the probability of a 1 GHz dark photon to have been converted to a standard model photon by today.

For a homogeneous universe, the resonance condition is met only at a few discrete redshifts as can be seen in Fig.~\ref{fig:mA}. But at $z\lesssim 6$, the universe is not homogeneous and we have matter over/under-densities. Therefore, at a given redshift, we still have a finite probability for the resonance condition to be satisfied, for a given $m_A$ which of course depends on sky location. Consequently, we need to carry out a continuous integral over redshift as in equation above. We discuss this point in detail in a  subsequent section. 

\vspace{-5mm}
\section{Aspects of the model}
\label{sec:main}
In this section, we go through the ingredients of C22 model in detail. We explain the origin of deviation from a pure power law, with the goal to clarify the different effects. 
We start with a brief summary of the existing modeling and data, and then add various effects in a step-by-step manner.

\vspace{-4mm}
\subsection{Considered data sets and previous models from C22}
\label{subsec:data}

\begin{figure}
\centering 
\includegraphics[width=0.95\columnwidth]{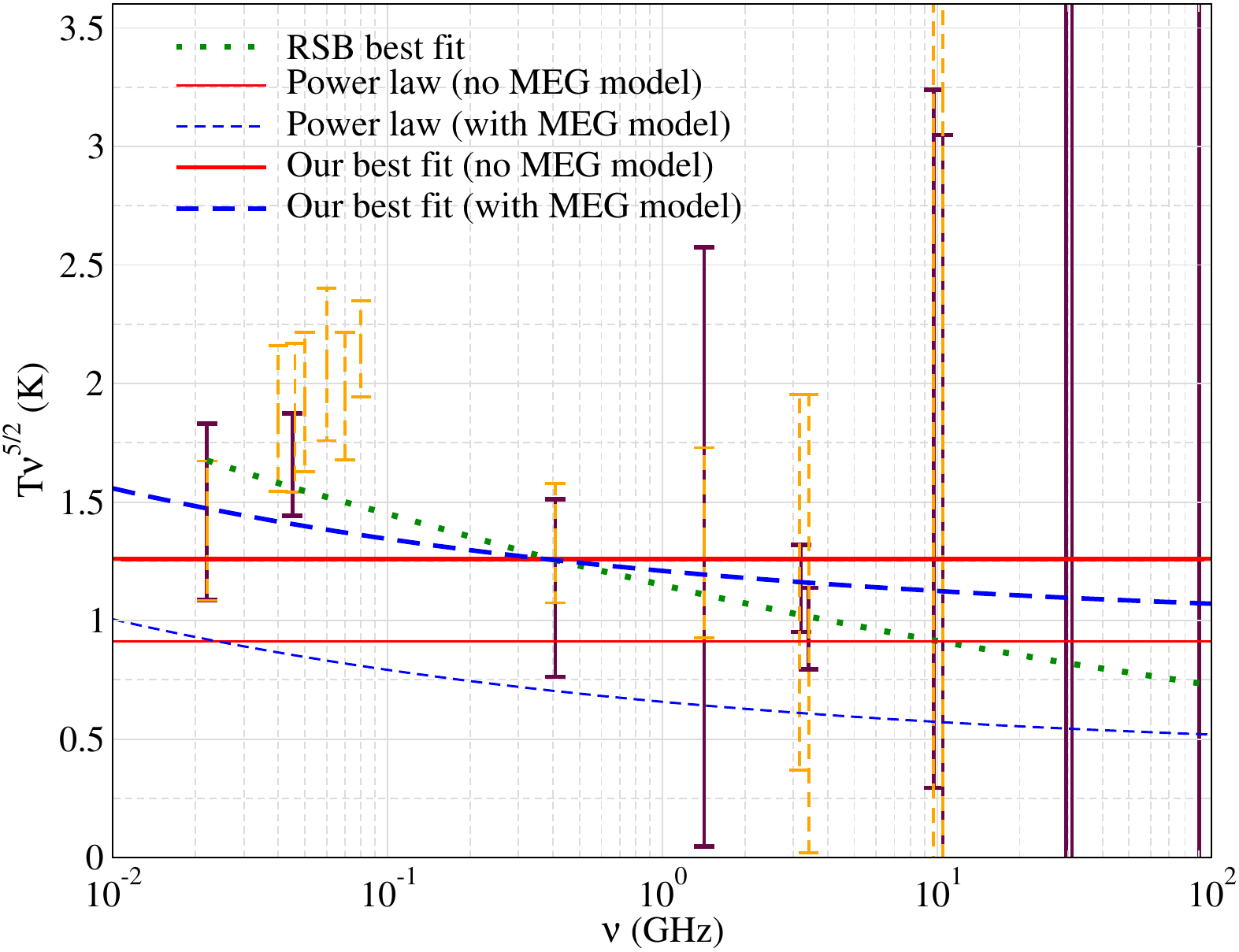}
\caption{Excess radio background and comparison with RSB  \citep{ARCADE2011} best-fit and data points (solid maroon). We scaled out the factor of $\nu^{-5/2}$ for clarity. We also show the full set of data points of \citet{DT2018} in orange, which includes re-analyzed ARCADE data points and new data points at 40-80 MHz. For the pure power-law model, we used the median values of posterior distribution of C22 (Table \ref{tab:table1}) which are shown as thin lines. Our refitted model is shown as thick lines. In this figure, we assume a simplistic pure power-law spectrum of excess radio background.}
\label{fig:caputo}
\vspace{-3mm}
\end{figure}

In Fig.~\ref{fig:caputo}, we present the brightness temperature for the RSB best-fit \citep{ARCADE2011} and the data compiled in Table 4 of the same paper. The ARCADE data points are between 3-90 GHz while the remaining data points are compiled from other experiments. 
For comparison, we also show the full set of data points of \cite{DT2018} in Fig.~\ref{fig:caputo}. The slope of the best-fit to this data set is consistent with the ARCADE analysis, but with slightly higher normalization (30 K at 310 MHz as opposed to 24 K for \cite{ARCADE2011}). 

\begin{table}
  \begin{center}
   \begin{tabular}{l|c|c|c|c} 
     & ${\rm log}_{10}m_X$ 
     & ${\rm log}_{10}m_{A}$ 
     & ${\rm log}_{10}(\tau_X)$ 
     & ${\rm log}_{10}(\epsilon)$ 
     \\
     & [eV] & [eV] & [s] &  \\
    \hline
    w MEG & $2.2\times 10^{-4}$ & $3.0\times 10^{-14}$ & $1.4\times 10^{21}$ & $1.1\times 10^{-7}$ \\
    w/o MEG & $1.1\times 10^{-4}$ & $6.0\times 10^{-14}$ & $3.0\times 10^{21}$ & $5.9\times 10^{-8}$
    \end{tabular}
  \end{center}
  \caption{Median values of posterior in Table I of C22. In both cases, C22 assume $T_{A,0}/T_0\approx 0.22$. This implies $P_{\rm GHz}=1.05\times 10^{-5}$ without EG model  and $P_{\rm GHz}=1.3\times 10^{-5}$ with EG model.}
  \label{tab:table1}
    \vspace{-2mm}
\end{table}

In the baseline model of C22, an irreducible extra-galactic  component was added to the predicted signal from their model. We use this as the minimal extra-galactic background (MEG). The fitting function to this background is given by \citep{GTZBS2008},
\begin{equation}
    T_{\rm bg}(\nu)=0.23\,{\rm K}\left(\frac{\nu}{\rm GHz}\right)^{-2.7}. 
    \label{eq:extra-galactic}
\end{equation}
C22 found that their baseline model provides a good fit to both data sets, even if slightly worse than a single power-law fit with a slope $\gamma\approx -2.6$, as obtained by \cite{ARCADE2011}. 
C22 also carried out a consistency test without the MEG component, still finding good agreement with the data. 

We show our version of the C22 best fit with the median parameters (Table \ref{tab:table1}), but assuming the model to be a pure power law as in Eq.~\eqref{eq:temp_amplitude}. We take $T_0$ to be 2.725~K \citep{Fixsen1996}, which was the median value obtained in C22, and do not vary this parameter in what follows. We find a small offset between the RSD data and the model, which might be due to the factor $P_{A\rightarrow \gamma}$ and its detailed modelling. We also show our best fit in Fig. \ref{fig:caputo}. To obtain the best fit, we had to tune $P_{\rm GHz}$ such that $P_{{\rm GHz}}=1.45\times 10^{-5}$ without EG model  and $P_{\rm GHz}=3\times 10^{-5}$ with EG model. From Eq.~\eqref{eq:dP_dz}, one finds that $P_{\rm GHz}\propto\epsilon^2$. Therefore, we need to increase $\epsilon$ by a factor of $\lesssim$1.5 from C22 value in Table \ref{tab:table1} in order to improve the fit. This is within 1-$\sigma$ uncertainty of the median values of C22. Alternatively, we could tune other parameters as this just rescales the spectrum in Eq.~\eqref{eq:temp_amplitude}.
For the discussion given below, we will use the power-law model with the C22 median parameters without MEG from Table~\ref{tab:table1} as a baseline to illustrate the importance of various effects.

\vspace{-3mm}
\subsection{Dependence on the Hubble parameter} 

\begin{figure}
\centering 
\includegraphics[width=0.95\columnwidth]{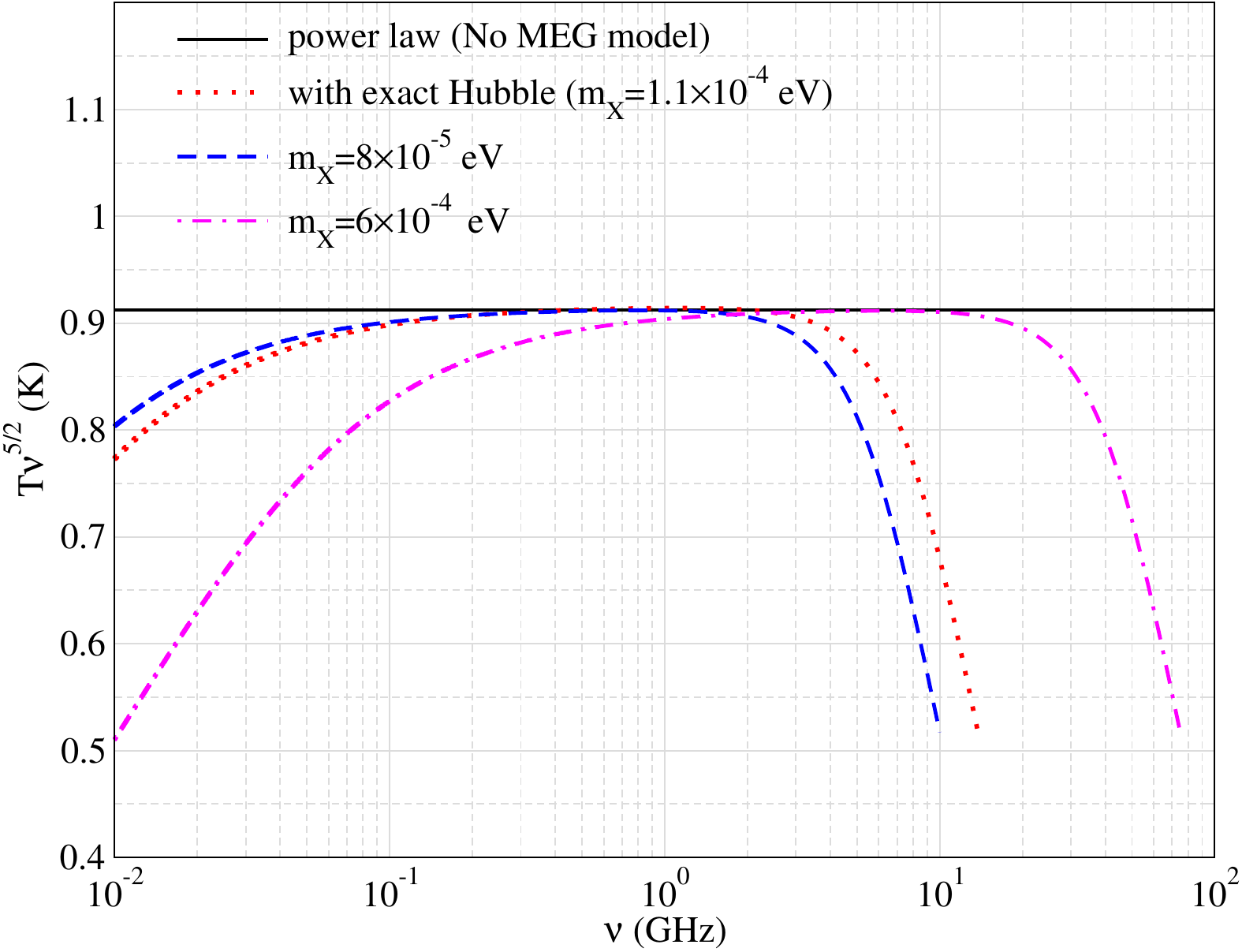}
\caption{Comparison of the power-law model with using exact Hubble. We have used the median of C22 (Table \ref{tab:table1}) as our fiducial values. We have ignored MEG radio background contribution. We have plotted two other cases with upper/lower limit of mass to showcase the importance of exact Hubble. We have rescaled these two cases so that the amplitudes of all cases match. }
\label{fig:arcade_no_stim}
\vspace{-3mm}
\end{figure}
In this section, we study the importance of using the exact expression for the Hubble parameter in the calculation. 
In Fig. \ref{fig:arcade_no_stim}, we compare our baseline power-law spectrum with the more accurate calculation for the fiducial case, $m_X=1.1\times 10^{-4}$~eV (Table \ref{tab:table1}). One can clearly see the importance of energy density contributions from radiation and dark energy at the low- and high-frequency ends, respectively, with significant deviations which should worsen the fit to RSB data.  We also choose two extreme cases with $m_X=8\times 10^{-5}$eV with $z_{10 {\rm GHz}}\approx 0$ (see Appendix~\ref{sec:appendix} for details) and $6\times 10^{-4}$ with $z_{10 {\rm GHz}}\approx 6$. This choice of parameters implies that the dark photons corresponding to photons, which show up today at 10 GHz, were emitted at $z=0$ or 6, respectively. This also means that dark photons corresponding to $\approx 20$ MHz photons were produced around $z\approx 455$ or 3400, i.e. $z_{20 {\rm MHz}} \approx 455$ or 3400 respectively. The later case should be the most sensitive to radiation energy density which one clearly sees in the figure.  

For the case with $m_X=1.1\times 10^{-4}$ eV, $z_{10 {\rm GHz}}\approx 0.25$ and $z_{20 {\rm MHz}}\approx 600$. At $z\simeq 600$, the radiation energy density is smaller than the matter energy density by a factor of $\approx 600/3400\simeq 0.18$. Therefore, the Hubble constant is $\simeq 10$ percent larger when compared to the case when we ignore radiation. For the other case with $m_X=6\times 10^{-4}$, the difference at 22 MHz turns out to be $\approx 40$ percent. For cases with $z_{10 {\rm GHz}}$=1,2,3,4 and 5, the change in intensity due to correct Hubble factor at 22 MHz turns out to be 14, 20, 26, 32 and 37 percent respectively. The uncertainty in the data point at 22 MHz is $\approx$ 25 percent \citep{ARCADE2011}. A simple criterion that the change in intensity at 22 MHz should be less than the uncertainty in the data point tells us that $z_{10 {\rm GHz}}\lesssim$3.  This then means that the $z_{20 {\rm MHz}}=2000$ as opposed to $z_{\rm eq}\approx 3400$ which was used to obtain the allowed parameter range of $m_X$ in the previous section. Our calculations would suggest that the parameter space is tightened by a factor of $\approx 1.7$ i.e. $8\times 10^{-5}\lesssim m_X \lesssim 3.5\times 10^{-4}$ eV. We will quantify this statement more accurately in Sec.~\ref{sec:reanalysis}.

One also notices the presence of curvature around $\approx 10$ GHz and $70$ GHz for the cases with $m_X=8\times 10^{-5}$ eV and $6\times 10^{-4}$ eV respectively. This is due to the presence of dark energy. The dark photons which are emitted at $z\lesssim$ 1 are the most affected due to the presence of dark energy, which leads to accelerated expansion and hence dilution of the dark photons. Our discussion proves that, at least from the theoretical point of view, one expects significant deviation from a perfect power law between $\approx$20 MHz and 10 GHz.

\vspace{-2mm}
\subsection{Relaxing soft photon limit for the stimulated decay}

\begin{figure}
\centering 
\includegraphics[width=0.95\columnwidth]{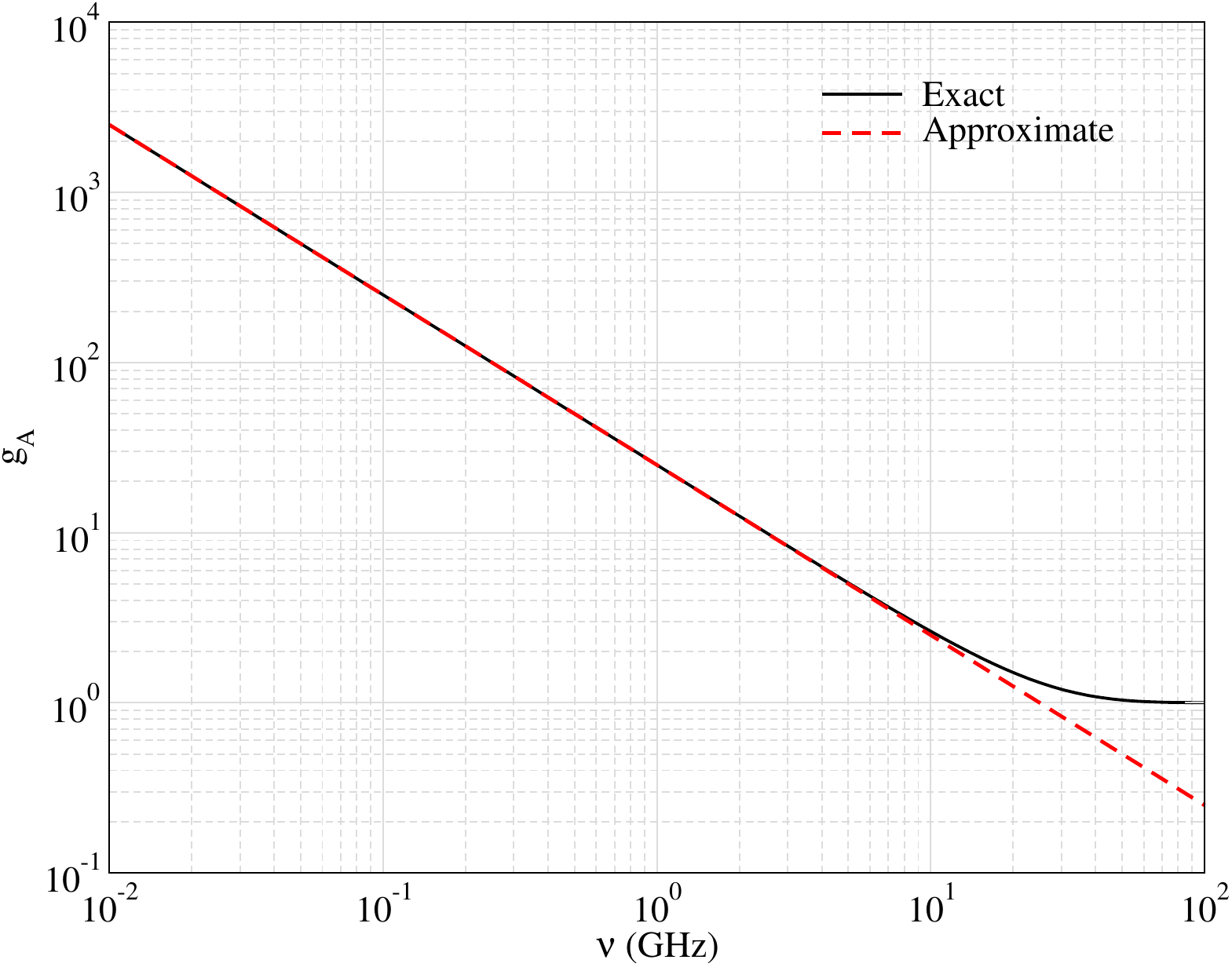}
\caption{Comparison of stimulation term $g_A$ and the approximation $\frac{0.44}{x}$ where $x=\nu/T_0$, where $\nu$ is the frequency of photons as seen today. We have used $T_{A,0}=0.22T_0$ which is the median of parameter space of C22.}
\label{fig:stimulated_term}
\vspace{-3mm}
\end{figure}

\begin{figure}
\centering 
\includegraphics[width=\columnwidth]{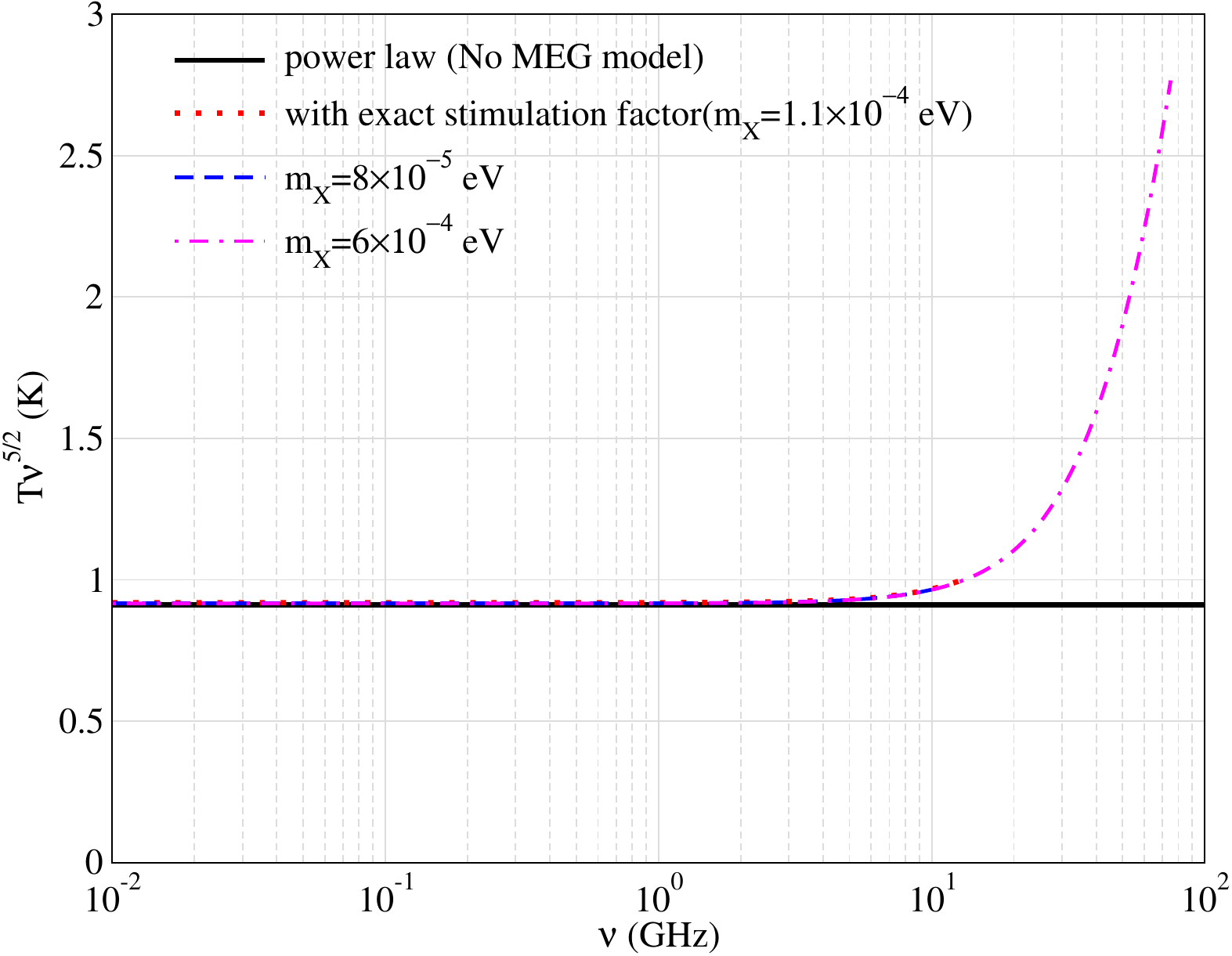}
\\
\caption{Comparison of power-law model with the calculation using the exact stimulation factor.}
\label{fig:arcade_stim}
\vspace{-3mm}
\end{figure}

We now discuss the modeling of the stimulated decay. We remind the reader that in C22 the decay lifetime of dark matter to dark photons depends on the occupation number of a thermal dark photon ($A$) background. The stimulation factor is given by,
\begin{equation}
    g_{A}(\nu_A)=1+2f_A=1+\frac{2}{{\rm e}^{\nu_A/T_A}-1},
\end{equation}
where $T_A$ is the dark photon temperature. For a standard photon that at $z=0$ is received at frequency $\nu$, the dark matter particle decayed at $1+z_*=\nu_{A,0}/\nu$. At this redshift, the stimulation factor therefore has to be evaluated at $x_A(\equiv\nu_A/T_A)= \nu_{A,0}/T_A(z_*)=\nu/T_{A,0}=x\,T_0/T_{A,0}$.
In Fig.~\ref{fig:stimulated_term}, we compare the soft photon approximation ($\nu_A\ll T_A$), which was also used in the simple power-law approximation, Eq.~\eqref{eq:temp_amplitude}, to the exact term. One can see clear deviations in high frequency regime, where the stimulated effects become negligible. 
 
In Fig.~\ref{fig:arcade_stim}, we plot the radio brightness for the cases considered in Fig.~\ref{fig:arcade_no_stim} but with the exact stimulation factor taken into account. We can see the effect of this term in terms of adding more curvature in the high frequency end, returning back to the brightness without stimulation, i.e., $T\propto (\nu/\nu_0)^{-3/2}$. We see that the modification is independent of $m_X$, at least for $\nu$ in the 20~MHz-10~GHz range.
This is because when increasing $m_X$, the dark photon had to be emitted at a higher redshift. Therefore, after the conversion the photons show up at the same frequency, $\nu$, independent of the dark matter mass.

\vspace{-3mm}
\subsection{Importance of inhomogenities in electron number density}
The electron number density below $z\lesssim 6$ is inhomogeneous and assuming it to be homogeneous as in Fig. \ref{fig:mA} is an over-simplification. The effect of inhomogeneity in electron density was considered in \cite{CLMR2020}. For an homogeneous number density of electrons, the resonance condition is met only at a few discrete redshifts. But in an inhomogeneous universe, the resonance condition can be met over a range of redshifts depending on sky location. One can then take a sky average to compute the radio background, which modified the overall dark photon to photon conversion probability. 

The differential probability of converting a dark photon to a photon is given by \citep{CLMR2020},
\begin{equation}
    \frac{{\rm d}P_{A\rightarrow \gamma}}{{\rm d}z}(z)=\frac{\pi m^4_A\epsilon^2}{\nu_A(t)}\left|\frac{{\rm d}t}{{\rm d}z}\right|f(m^2_{\gamma}=m^2_A,t),
    \label{eq:dP_dz}
\end{equation}
with $f(m^2_{\gamma},t)=\frac{\mathcal{P}(\delta (m^2_{\gamma},t))}{\bar{m}^{*2}_{\gamma}}$, where $\delta=\frac{m^2_{\gamma}}{\bar{m}^{*2}_{\gamma}}-1$ and $\mathcal{P}(\delta,z)$ is the probability distribution function with,
\begin{equation}
    \int \mathcal{P}(\delta,z) \id\delta=1
\end{equation}
The authors in \cite{CLMR2020} considered two types PDFs, one gaussian and another log-normal. The gaussian PDF is given by,
\begin{equation}
    \mathcal{P}_G(\delta,z)=\frac{1}{\sqrt{2\pi\sigma^2(z)}}{\rm exp}\left(-\frac{\delta^2}{2\sigma^2(z)}\right).
\end{equation}
Similarly, the log-normal PDF is given by,
\begin{equation}
    \mathcal{P}_{LN}(\delta,z)=\frac{(1+\delta)^{-1}}{\sqrt{2\pi\Sigma^2(z)}}{\rm exp}\left(-\frac{{\rm ln}(1+\delta)+\Sigma^2(z)/2}{2\Sigma^2(z)}\right),
\end{equation}
where $\Sigma^2(z)={\rm ln}(1+\sigma^2(z))$ and $\sigma(z)$ is the variance of baryon number density fluctuations. In linear regime, $\sigma(z)\ll 1$ and both PDFs give similar results. But once $\sigma(z)\gtrsim 1$, the gaussian PDF becomes unphysical as it assigns finite probability to $\delta< -1$ while log-normal PDF is constrained to be always valid for $\delta> -1$.

\begin{figure}
\centering 
\includegraphics[width=\columnwidth]{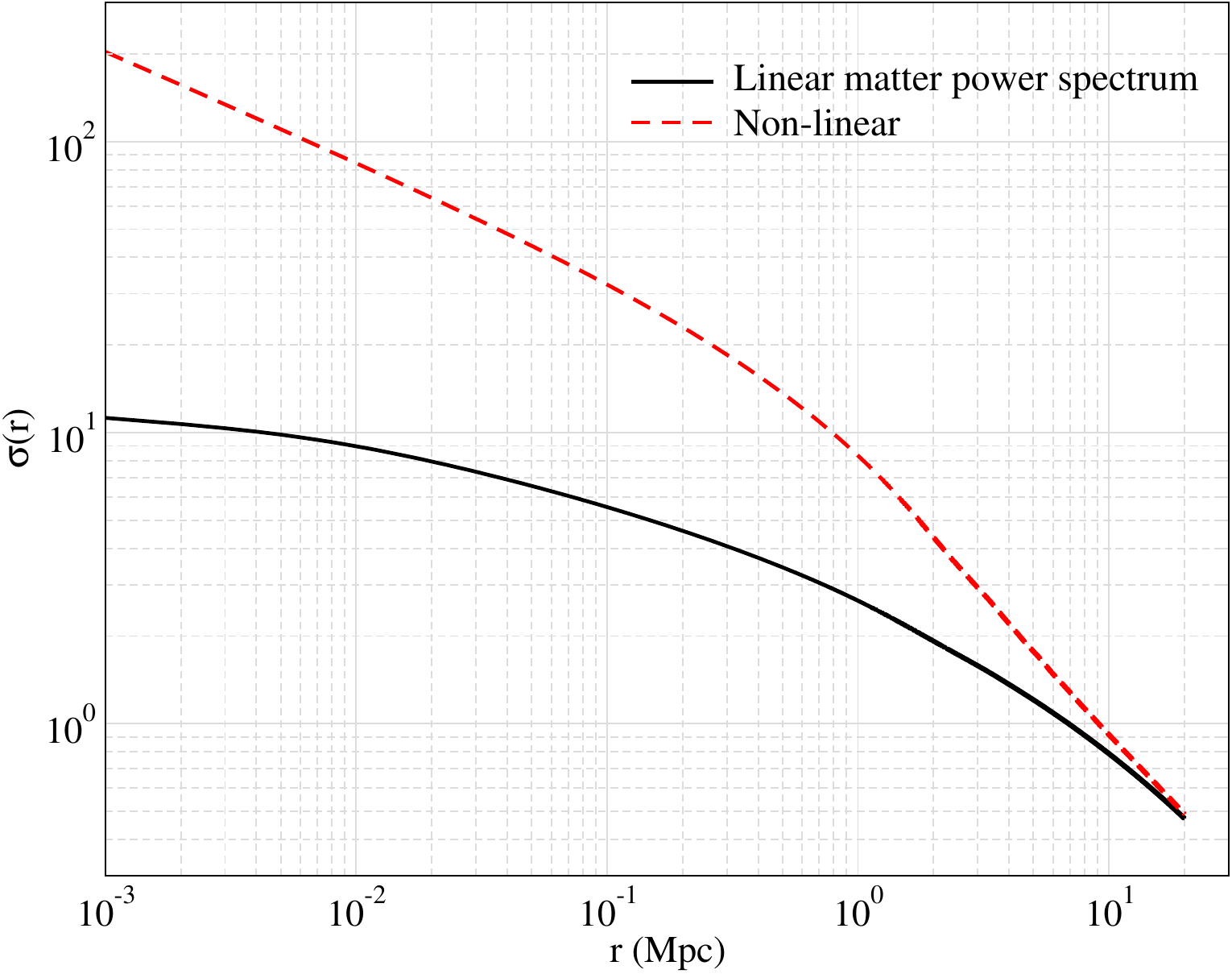}
\\
\caption{Variance of mass fluctuations using linear and non-linear matter power spectrum from {\tt CLASS} \citep{CLASSCODE}. }
\label{fig:variance}
\end{figure}
The variance of fluctuations at a scale $r$ is computed by smoothing the density power spectrum with a smoothing scale $r$. The expression is given by,
\begin{equation}
    \sigma(r)=\int \frac{\id^3k}{(2\pi)^3}j_0(kr)P_b(k,z)
\end{equation}
where $P_b(k,z)$ is the baryon density fluctuations. Since dark photon-photon conversion is local in space, we need small scale information which are highly non-linear and typically need simulations. Here we ignore such complications and assume that baryonic density power spectrum is given by the matter power spectrum. We can then use the power spectrum obtained from \cite{CLASSCODE} and compute the variance as a function of length scale. We show this result in Fig.~\ref{fig:variance}. We use the power spectrum up to $k=10^4$ Mpc$^{-1}$ and we have checked that we can compute variance to length of $\sim$ kpc reliably. We choose this as the cutoff length scale below which there is no baryon density perturbation. In that case, $\sigma(z=0)\simeq 10$ and 200 for linear and non-linear matter power spectrum respectively. This choice is on equal footing with the calculations of \cite{CLMR2020} which was used in C22. However, at such small length scales, pressure smoothing may already damp out small-scale structures \citep{KHORS2015}. Therefore, details of these calculations will be sensitive to small scale physics and is subject to further discussions. 

Our computed variance is of similar order to that of \cite{CLMR2020} (see Fig. 8 of the reference) who use simulations to compute the baryon density power spectrum. For the non-linear case, $\sigma(z=0)$ varies between 300-1000 \citep{CLMR2020}. In this work, we use a conservative value of $\sigma(z=0)=500$ and $\sigma(z)=\frac{\sigma(0)}{(1+z)}$ which captures the redshift evolution pretty well. C22 use the log-normal distribution to compute the probability, which we follow here.

\begin{figure}
\centering 
\includegraphics[width=\columnwidth]{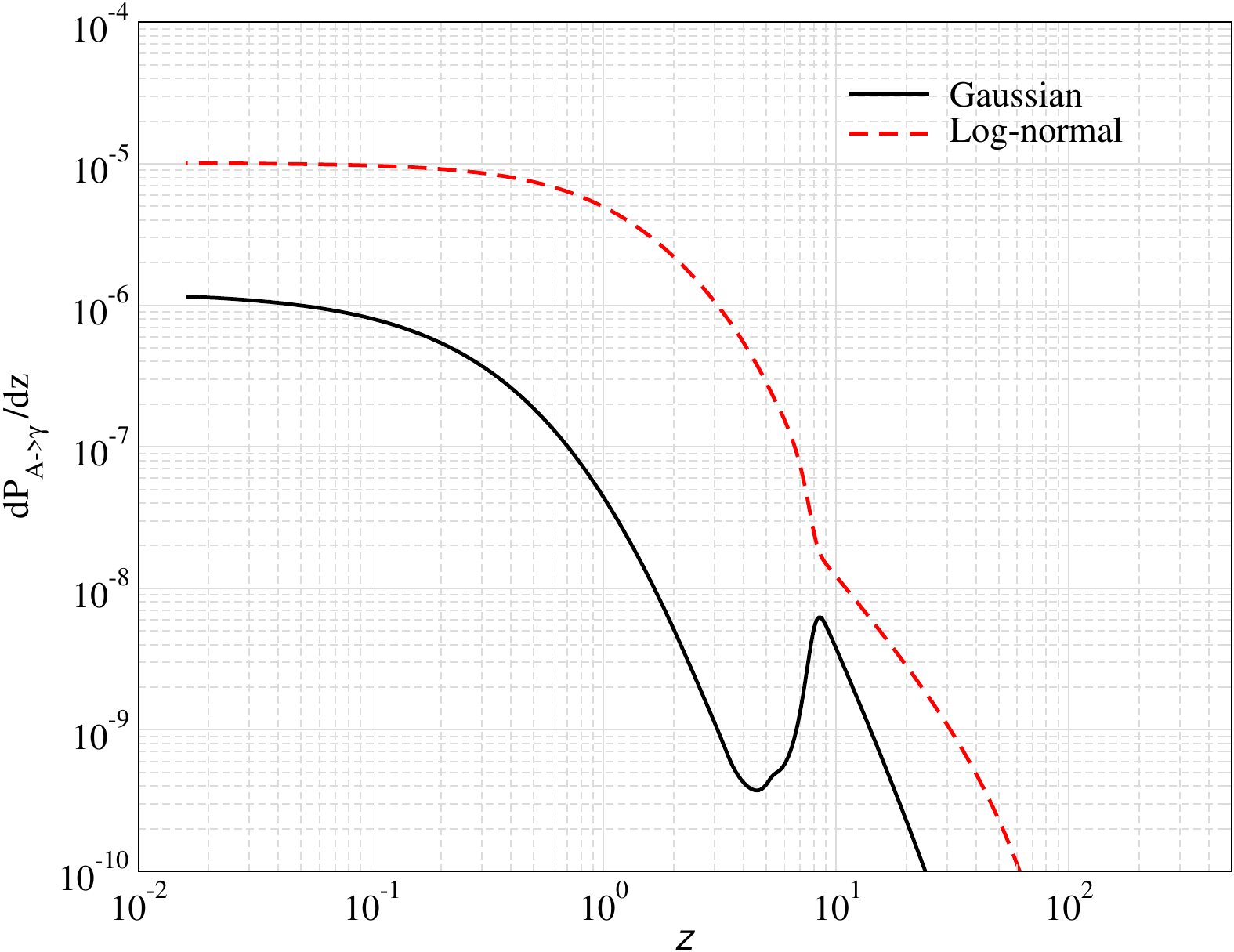}
\\
\caption{$\frac{{\rm d}P_{A\rightarrow \gamma}}{{\rm d}z}(z)$ with $m_A=3\times 10^{-14}$ eV, $\epsilon=10^{-7}$. }
\label{fig:dP_dz}
\end{figure}

\begin{figure}
\centering 
\includegraphics[width=\columnwidth]{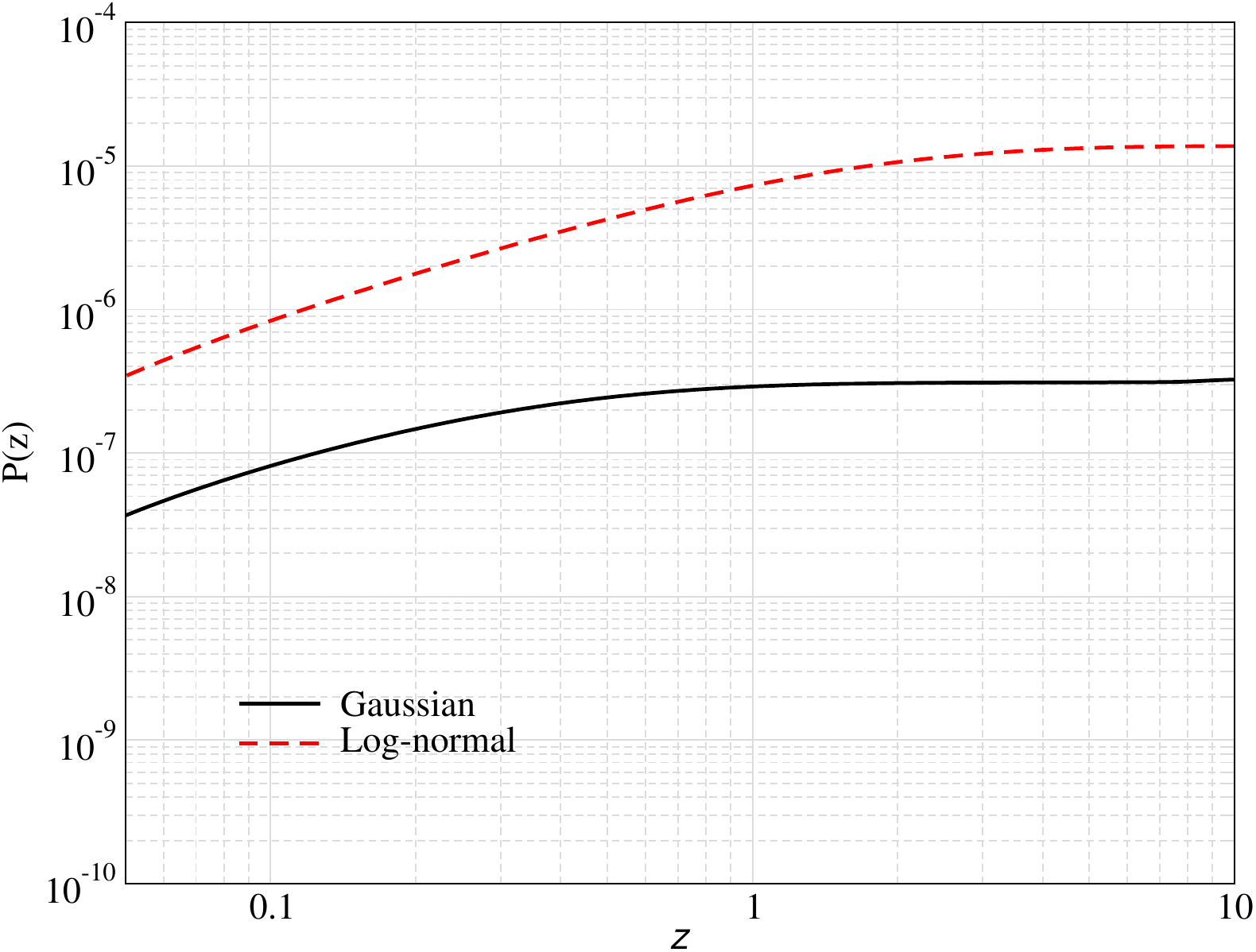}
\\
\caption{Probability that a dark photon which was emitted at $z$ is converted to photon by today with $m_A=3\times 10^{-14}$ eV, $\epsilon=10^{-7}$. We use the log-normal distribution in this work which the authors in C22 did as well. }
\label{fig:P_z}
\end{figure}

In Fig. \ref{fig:dP_dz}, we plot the redshift-differential probability of converting dark photon to photon with $m_A=3\times 10^{-14}$ eV. We see that most of the conversion is happening at $z\lesssim 6$. In Fig. \ref{fig:P_z}, we plot the dark photon conversion probability from the redshift of injection until today. We observe an overall enhancement for the non-Gaussian fluctuations.
From previous discussions it clear that dark photons corresponding to 10 GHz photons have to be emitted deep in the matter era ($z_{10 {\rm GHz}}\approx 0$) to have an approximate power law at low frequency. In that case the $z$ integral of $P_{A\rightarrow \gamma}$ has an implicit dependence on frequency due to the allowed range of integral of $z$. This leads to curvature of intensity of radio excess at high frequency end according to Fig.~\ref{fig:P_z}. 

\begin{figure}
\centering 
\includegraphics[width=\columnwidth]{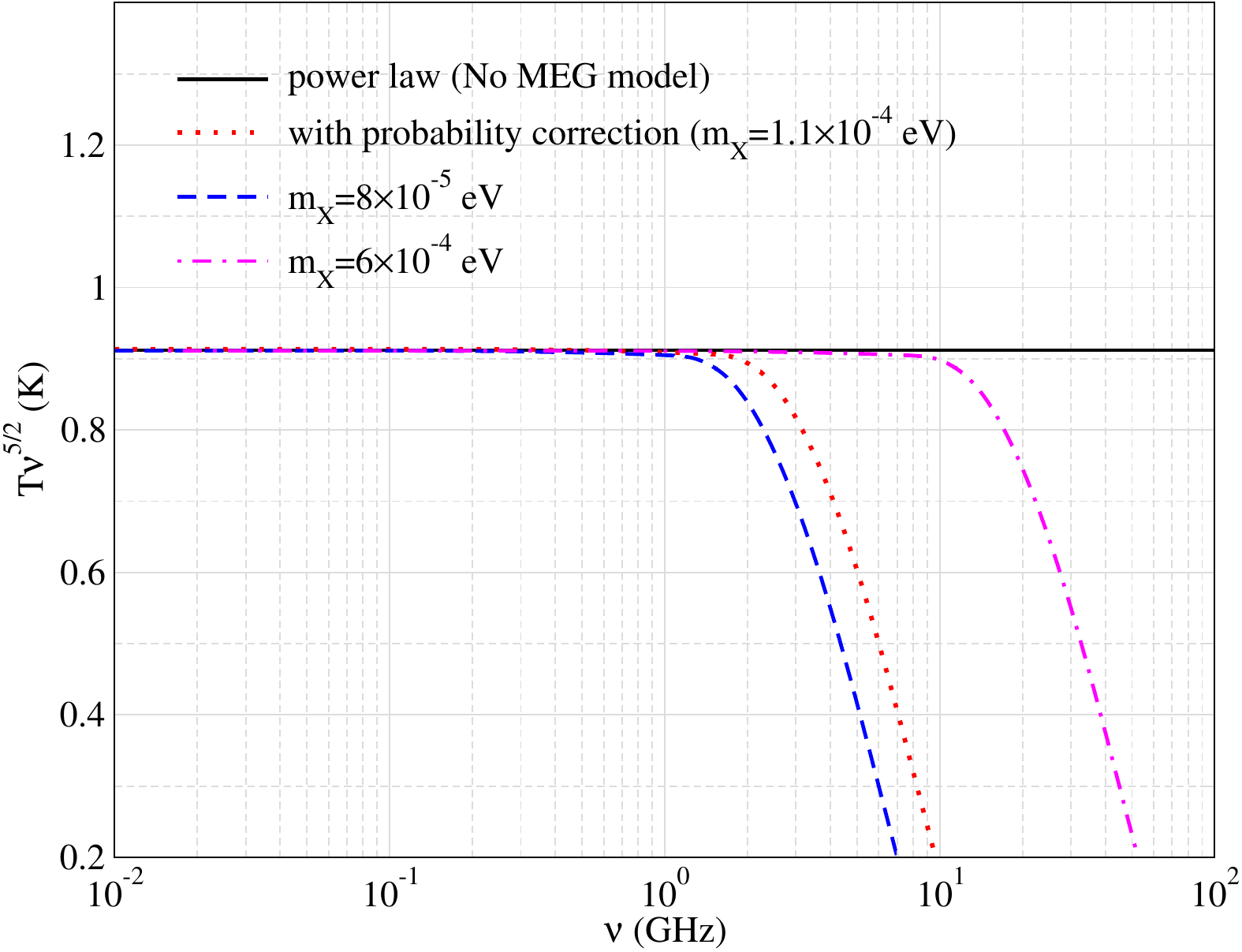}
\\
\caption{Comparison of power-law model with calculation using frequency dependent probability of conversion. }
\label{fig:P_z_correction}
\end{figure}

In Fig. \ref{fig:P_z_correction}, we show the importance of including this aspect for a few example. For our fiducial case with $m_X=1.1\times 10^{-4}$eV, $z_{10 {\rm GHz}}\approx 0.25$, therefore, should be very sensitive to this correction. Indeed, we find a big change to the radio brightness at frequency $\simeq 1-10$GHz. The uncertainty of the RSB measurements at these frequencies is also larger (see Fig.~\ref{fig:caputo}). Therefore, more precise data at $\simeq 1-10$GHz may allow placing tight constraints on the parameter space of the model. 

\begin{figure}
\centering 
\includegraphics[width=\columnwidth]{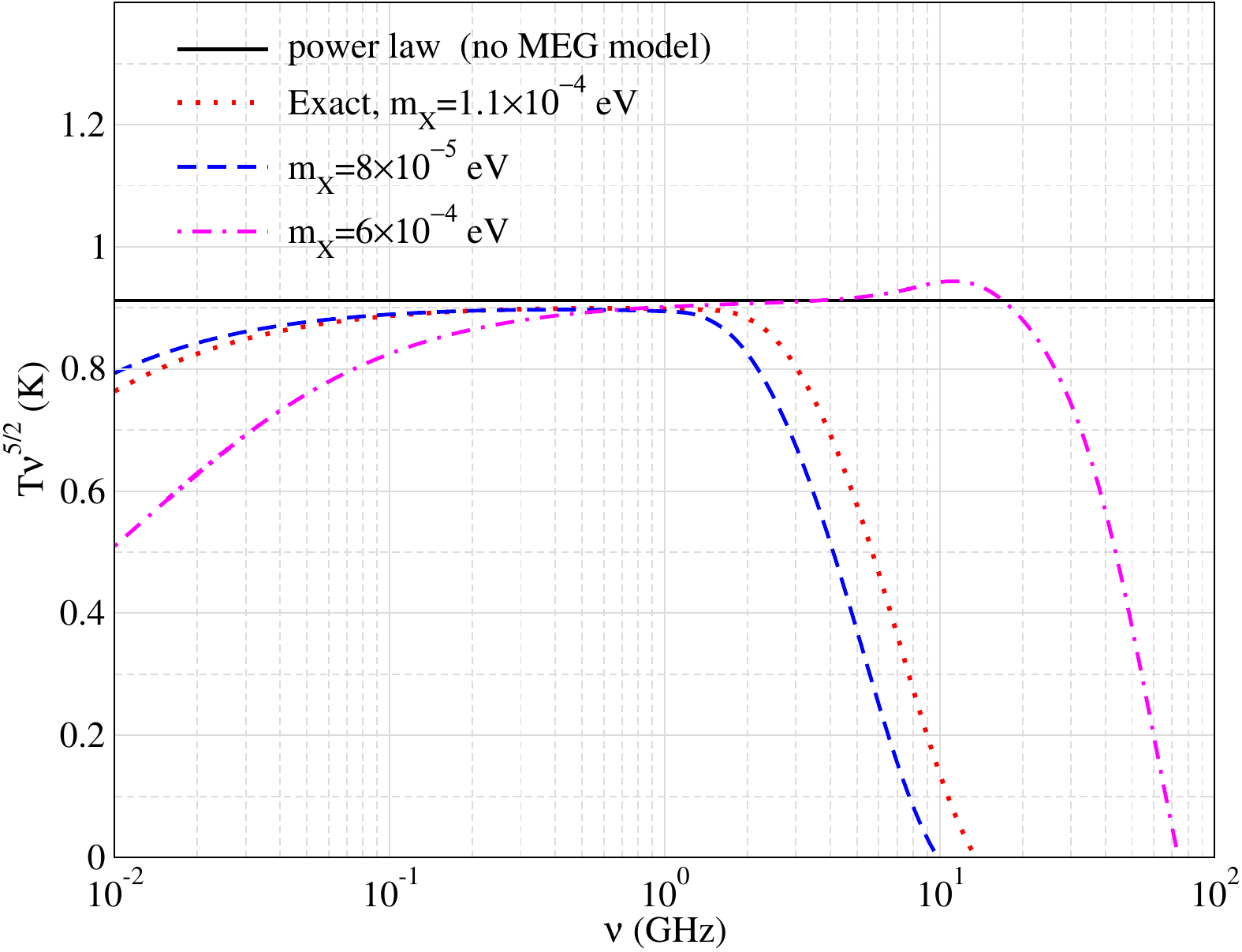}
\\
\caption{Comparison of power-law model with the exact calculations including all the discussed effects. Aside from the dark matter mass, none of the other model parameters have been varied. However, allowing for changes we can obtain an improved description of the data with the exact model.}
\label{fig:P_z_corrected_temp}
\end{figure}

\subsection{Cumulative effect}
Finally, in Fig.~\ref{fig:P_z_corrected_temp}, we show our calculations for the cumulative effect of all these corrections. There are noticeable departures from a simple power-law behaviour at both low and high frequencies. This showcases the fact that the assumption of pure or even approximate power-law spectrum is not justified, and hence must lead to biases parameter values. We will show below that the curved spectrum is actually provides a better representation of the RSB data when compared to the power-law model. 

\section{Simple reanalysis of the data}
\label{sec:reanalysis}

\begin{table*}
  \begin{center}
   \begin{tabular}{l|c|c|c|c|c|r} 
     & $m_X$ 
     & $m_{A}$ 
     & $(\tau_X)$ 
     & $\epsilon$
     & $T_{A,0}/T_0$
     \\
    & [eV] & [eV] & [s] &  &  & \\
    \hline
    prior & $(8-60)\times10^{-5}$ & $10^{-16}-10^{-11}$ & $10^{20}-10^{25}$ & $10^{-9}-10^{-6}$ & 0.05-0.4 \\
    With MEG  & $1.48\times 10^{-4}$ & $5.01\times 10^{-16}$ & $ 10^{20}$ & $10^{-6}$ & 0.22 \\
    No MEG  & $1.0\times 10^{-4}$ & $5.01\times 10^{-13}$ & $1.54\times 10^{24}$ & $10^{-6}$ & 0.11
    \end{tabular}
  \end{center}
  \caption{Prior and best fit parameters of our model with and without MEG background.}
  \label{tab:table2}
\end{table*}

We now discuss our simple method to find the best fit of our model to the RSB data. The parameters in our model are $m_X,m_A,\epsilon,\tau_X$ and $T_{A,0}/T_0$. Our choice of prior is given in Table~\ref{tab:table2} and is based on constraints on these parameters from other probes (a discussion can be found in C22). 

\begin{figure}
\centering 
\includegraphics[width=\columnwidth]{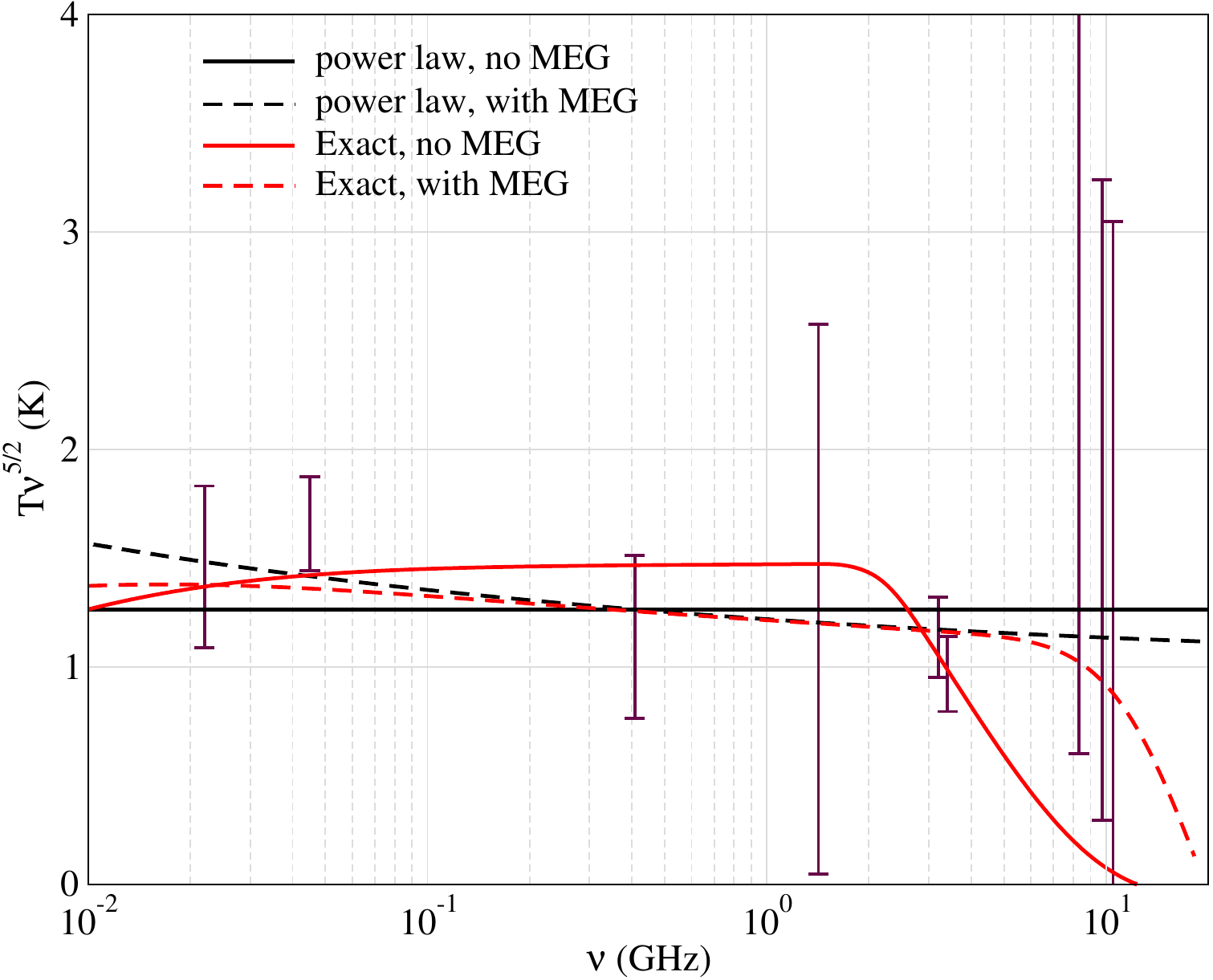}
\\
\caption{Best fit model for various scenarios as shown in the figure.}
\label{fig:chi2}
\end{figure}
We consider RSB data points from 22MHz-10.49 GHz \citep{ARCADE2011}, as shown in Fig.~\ref{fig:chi2}. To find the best fit, we carry out a simple $\chi^2$ minimization:
\begin{equation}
    \chi^2=\sum_i\frac{(O_i-E_i)^2}{\sigma^2},
\end{equation}
where $O_i$ is our model, and $E_i$ and $\sigma_i$ are the mean value and uncertainty of the data, respectively.

\begin{figure}
\centering 
\includegraphics[width=\columnwidth]{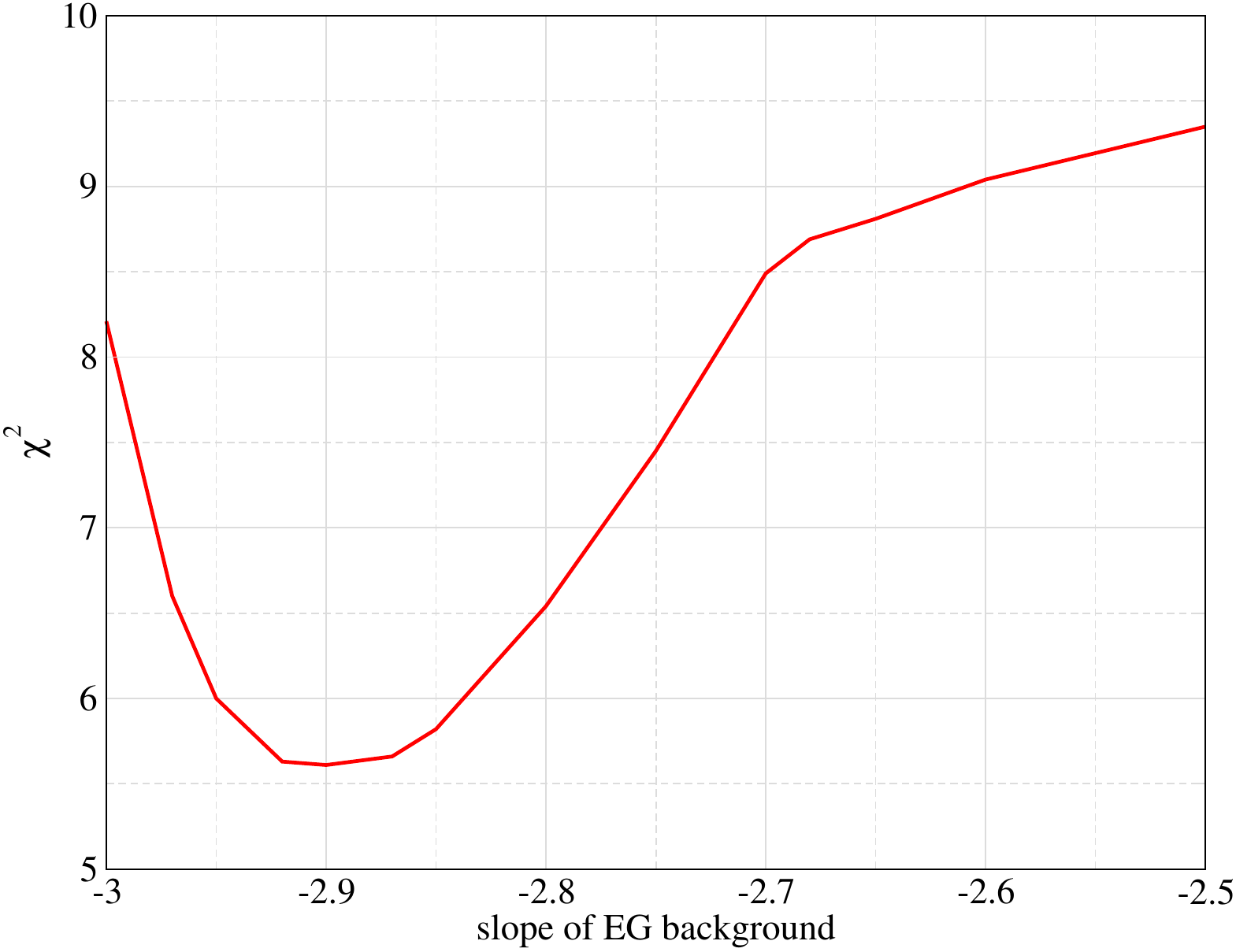}
\\
\caption{$\chi^2$ value by varying the slope of the MEG background. }
\label{fig:EG_bg_chi2}
\end{figure}

We first find the best fit for power-law model. Since the model is a pure power law, only the amplitude is the variable both with and without MEG model. 
It is easy to see that there is significant degeneracy between all the parameters in this case. The $\chi^2$ values for our best fit power-law model are 7.53 and 11.58 with MEG and without MEG model, respectively. This indicates that the power-law model without MEG provides a poorer description of the data.

Next we turn to the exact model. The best fit parameters are shown in
Table \ref{tab:table2}. The curved spectrum breaks a part of model degeneracy. There is still a residual degeneracy between $\epsilon$ and $\tau_X$ which are just scaling parameters and one can find a bigger best-fit parameter space by widening the priors. The curvature at low frequency end is entirely driven by $m_X$. To minimize the mismatch with low frequency data, one prefers $m_X$ such that dark photons are injected deep in matter era. We find the best fit $m_X$ to be $\approx 1\times 10^{-4}$ which corresponds to $z_{10 {\rm GHz}}=0.25$.
Including a MEG improve the overall agreement with the data. With the MEG model that we use, the spectrum becomes a better match at low frequency. The best fit $\chi^2$ values are 8.5 and 10.1 with and without MEG model respectively. Visually, the exact model looks like a better fit than the power law without MEG from Fig.~\ref{fig:chi2} except around highest frequency data points. With MEG, the high frequency part of our model looks like a slightly worse fit compared to power law which might be responsible for the slightly higher $\chi^2$ value. These discussions reiterates the importance of more and precise data at $1-10$ GHz to distinguish between particle-physics inspired models for the explanation of RSB excess. It also highlights the importance of using the full model to obtain the fit.   

We comment that the MEG model may not be robust at $\lesssim 100$ MHz, where an extrapolation was used. Since the MEG background is a steep power law, it becomes increasingly important at low frequencies. In Fig.~\ref{fig:EG_bg_chi2}, we vary the slope of the MEG background and obtain the minimum $\chi^2$ value. We see that a slightly steeper slope than what has been assumed here gives an improved fit, with $\chi^2 \simeq 6$. This compensates for downward curvature at low frequency from our model. If the slope becomes too steep, there are too many photons at low frequencies and the fit degrades. This illustrates the interplay between the assumptions of the MEG and the dark photon conversion model.


\section{Constraints from CMB spectral distortions}
\label{sec:spectral_distortion}


The conversion of dark photons to photons or vice versa does not need to happen only in the frequency range of interest considered in this work. In this section, we consider importance of this conversion in the frequency range 60-600 GHz. The distortion to CMB in this frequency band is well measured by {\it COBE/FIRAS} \citep{Fixsen1996}. The distortion to CMB spectrum in this frequency range is constrained to one part in $10^5$. Therefore, this measurement provides a stringent constraint on the parameter space of the model. Here, we are interested in the possibility of both thermal and non-thermal dark photons converting to CMB photons within 60-600 GHz band. 

\begin{figure}
\centering 
\includegraphics[width=\columnwidth]{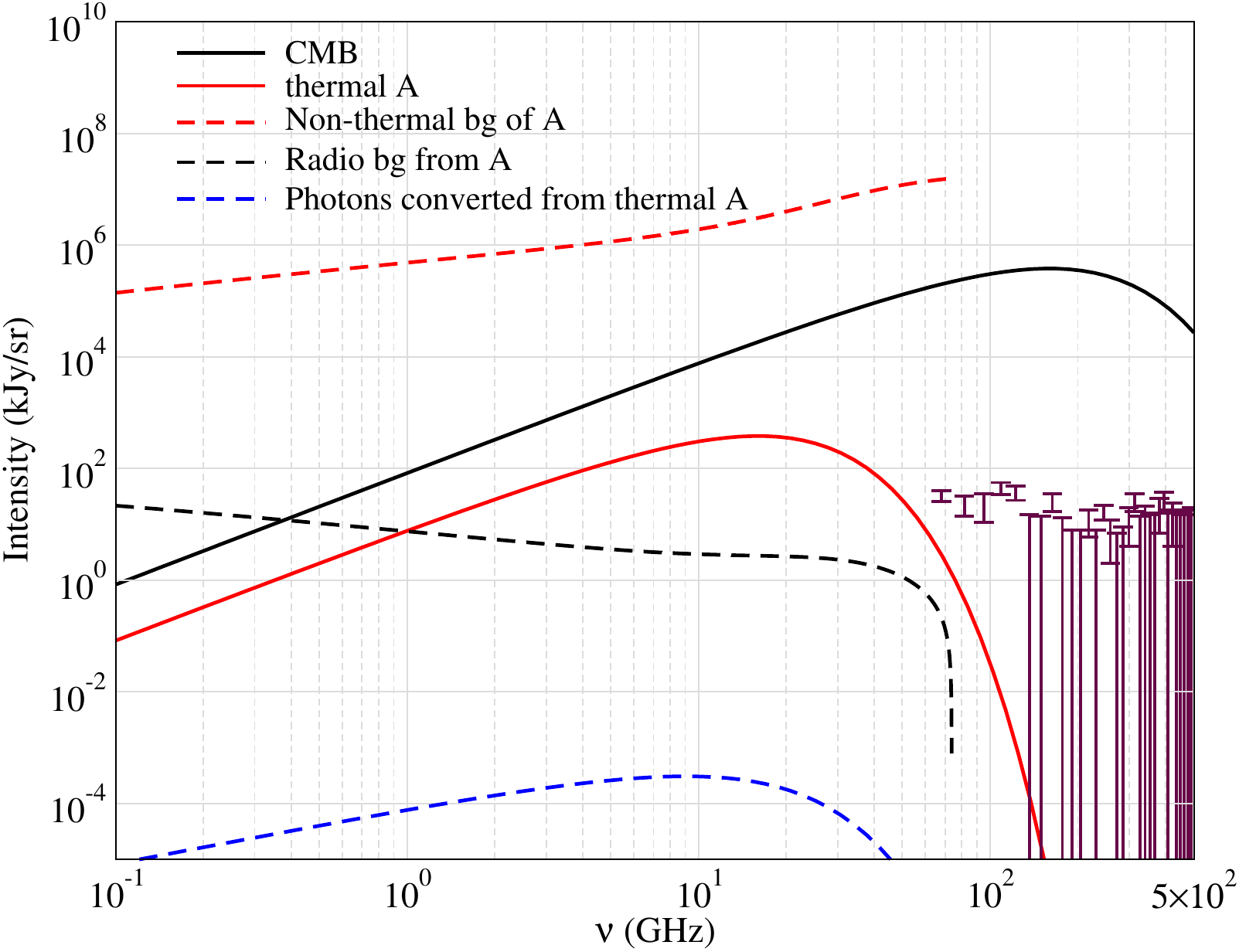}
\caption{Comparison of intensity of dark photon non-thermal, thermal background, radio excess from non-thermal dark photon background and CMB. We show the residuals and the uncertainty of data points from \citep{Fixsen1996} in solid maroon lines. We also predict CMB spectral distortion signal from resonant conversion of thermal dark photons. The parameters we use are, $m_X=6\times 10^{-4}$ eV $(z_{10 {\rm GHz}}=6)$, $\tau_X = 10^{20}$s, $T_{A,0}=0.1T_0$, $m_A=3\times 10^{-14}$ eV.  }
\label{fig:temperature_comp}
\vspace{-3mm}
\end{figure}

In Fig. \ref{fig:temperature_comp}, we compare the intensity of thermal/non-thermal dark photon background and the CMB for a particular parameter combination. We again remind the reader that in the model, the dark photons are assumed to be emitted in matter-dominated era. Therefore, in the extreme case, the lowest frequency photon ($\approx $ 20 MHz as seen today) has to be emitted at $z\approx 3400$. One can easily check that the rest frame energy of these dark photons would be $\approx 70$ GHz. Therefore, the most energetic photon (after converting dark photon) in this model is found at 70 GHz, which is just within {\it COBE/FIRAS} band. We see from Fig.~\ref{fig:temperature_comp} that the excess photons at $\approx 70$ GHz is about 2 order below {\it COBE/FIRAS} limit. However, this level of signal can in principle be probed by future spectral distortion mission such as {\it PIXIE} \citep{Kogut2011PIXIE} or {\it PRISM} \citep{PRISM2013WP, PRISM2013WPII}. 

Next we turn to the case of thermal dark photons getting converted to CMB photons. These thermal dark photons can themselves convert to photons when the resonance condition is met. The upper limit on temperature of dark photons is obtained by demanding that the total energy density of all relativistic species is not violated. In Fig. \ref{fig:temperature_comp}, we see that appreciable CMB spectral distortion can only be created in a very small band of 60-100 GHz beyond which the number density of the thermal dark photons becomes too small. In the mentioned frequency range, the intensity of the dark photon is roughly two orders of magnitudes smaller than the CMB. The probability of conversion $P_{A\rightarrow \gamma}$ would be of the order of $10^{-6}-10^{-7}$ at $\approx 10^2$ GHz for our choice of fiducial parameters. Therefore, we need to boost the probability by four orders of magnitude such that distortions in CMB is of the order of $\sim 10^{-5}$, and hence make the signal visible to \COBEF. Looking at the expression in Eq.~\eqref{eq:dP_dz}, one may believe that by boosting $m_A$ by an order of magnitude or a combination of $\epsilon$ and $m_A$ could boost the probability of conversion to sufficiently high values. But we want to remind the reader that conversion of efficiency is highest when $m_A=\bar{m}^*_{\gamma}$. Pushing $m_A$ to too high a value will thus have the opposite effect of driving the conversion efficiency down. More importantly, we note that the fiducial values do not violate the {\it COBE/FIRAS} constraints and we need to push the allowed parameters to extreme limit to see any visible signature. Therefore, we expect that CMB spectral distortions do not provide strong limits on the allowed parameter space, currently. But the predicted spectral distortion signal will be of the order of $10^{-8}-10^{-9}$ and future mission such as {\it PIXIE} can probe and put strong constraints on such models. Similar conclusions were reached by C22.

\subsection{Photon injection from decay in radiation era}
\begin{figure}
\centering 
\includegraphics[width=0.95\columnwidth]{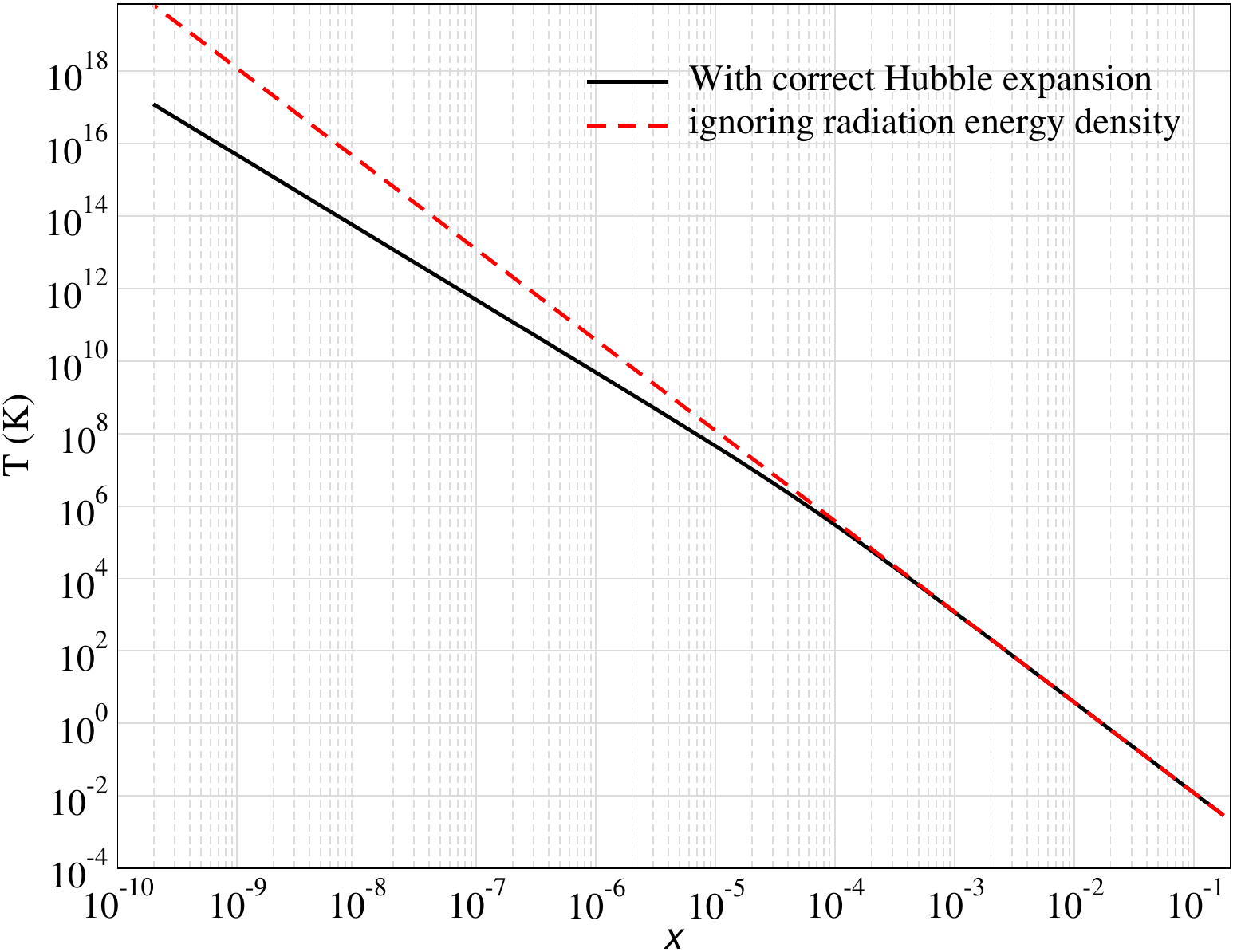}
\caption{Radio background from dark photon converting to photons at $z=0$ as a function of $x=\nu/T_0$. We have taken the dark photon emission from the radiation era into account.}
\label{fig:spectrum_radiation}
\vspace{-3mm}
\end{figure}

In the model, the dark matter keeps continually decaying throughout the history and it holds true even for radiation-dominated universe. It is easy to see that in the radiation era, $\frac{{\rm d}N_\gamma}{{\rm d}x}\propto x^{-1}$ and hence $T\propto x^{-2}$. In Fig. \ref{fig:spectrum_radiation}, we plot the radio spectrum from dark matter decaying to dark photons which then convert to photons at $z\lesssim 6$. Dark photons emitted at sufficiently high redshift will be non-relativistic today. To avoid this complication, we can choose the cut-off redshift such that the energy of dark photons today is equal to its mass, i.e. $\frac{\nu_{A,0}}{1+z_{\rm cut}}=m_A$.

Because the intensity is independent of frequency (as can be seen above), the total energy density is dominated by dark photons at the highest frequency. Therefore, even if, the non-relativistic dark photons can be taken into account, the energy density will be low enough to have any interesting effect. Once the dark photon is converted, a fraction of the ultra-soft photons can be absorbed by the electrons via thermal free-free absorption, which can lead to heating and therefore $y$-distortions \citep{Chluba2015GreensII}. Assuming that soft photons below $x\lesssim 10^{-4}$ are all converted into heat, the total energy density which can be potentially turned into heat is around $10^{-9}$ eV/cm$^{-3}$. Beyond $x\approx 10^{-4}$, the photons show up as radio excess today ($20$ MHz corresponds to $x\approx 3\times 10^{-4}$). At $z\lesssim 6$, the temperature of the gas is $\sim 10^4$K. The baryon energy density today turns out to be $\sim 10^{-7}$ eV/cm$^{-3}$. Therefore, the heat available from absorbed soft photons should not affect the evolution of gas temperature at $z\lesssim 6$ by more than a few percent and hence remains unobservable.

\section{Limitation of the model}
\label{sec:limitations}

Up to this point, we have discussed the various aspects of the model considered in C22. In this section, some of the possible challenges to this model are presented. 

\subsection{Anisotropies of radio background}
In previous sections, we only dealt with the sky-averaged radio excess signal. But from the discussions on inhomogeneous electron distribution, it is clear that this radio excess is anisotropic. Anisotropies in radio background depends upon two factors, (1) anisotropies of source which is the dark matter in this case and (2) anisotropies of scatterers which is the electron number density. The authors in C22 claimed that the anisotropies in source distribution can be neglected as the dark photons are emitted predominantly at $z\gtrsim 5$. We stress that in order for this statement to be true, dark photons corresponding to 10 GHz photons have to be emitted at redshifts larger than 5 i.e. $z_{10 {\rm GHz}}\gtrsim 5$. Considering, only the contribution from electron number density, the predicted radio anisotropy seems to be in agreement with current data (Fig. 3 of C22). 

However, as we saw in the previous section, our best fit values as well as the median of posterior distribution of C22 implies $z_{10 {\rm GHz}}$ which are well within $z=5$. Then, one has to take into account the anisotropy in the dark matter distribution. At $z\lesssim 5$, the dark matter is correlated with large scale structure. The clustered dark matter will then lead to anisotropies in radio background which is one order higher than measured anisotropies at 4-8 GHz \citep{Holder2014}. Therefore, even if we found a parameter space for the average background in this model, we still run into problems explaining the smoothness of radio background. Recent results show that measurements of radio background anisotropy are complicated and still subject of large debate \citep{OSHHL2022}. Therefore, future radio anisotropy measurement can place strong constraints on this model, potentially even ruling it out.
 
\subsection{Non-zero width of dark matter decay to dark radiation}
\label{sec:stim}

\begin{figure}
\centering 
\includegraphics[width=\columnwidth]{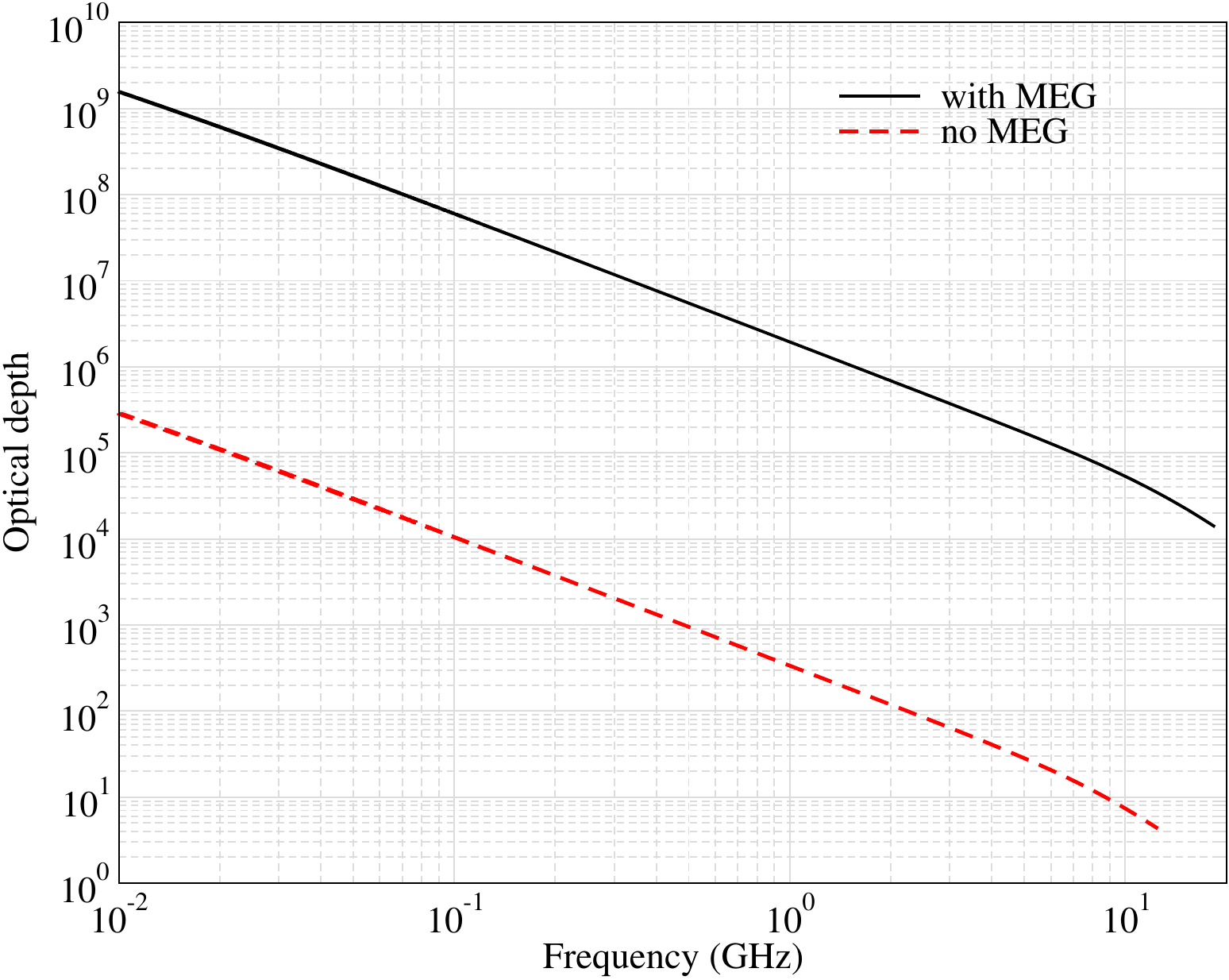}
\\
\caption{Optical depth ($\tau_x$) as a function of dark photon frequency today for the best fit parameters for the exact model cases as shown in Fig. \ref{fig:chi2}, which were obtained using the monchromatic injection approximation.}
\label{fig:optical_depth}
\end{figure}

One of the requirement for the approximate power law spectrum is to have a thermal distribution of background dark photons during the stimulated decay of dark matter. The authors in C22 discussed the possibility of distortion in the thermal spectrum of dark photons in their Appendix B. The authors showed that within their model, it is likely that the distortion is negligible. However, in the calculations, it was assumed that the decay product is monochromatic. In this section, we consider a new effect due to non-zero width of decay.

One of the key aspect of the model is the boost in decay rate of dark matter due to stimulation by thermal dark radiation background. In principle, the total dark radiation background (both thermal and non-thermal) would lead to stimulation. But with the assumption of a monochromatic injection spectrum of dark radiation as the decay product, the non-thermal component does not contribute: since there is a one-to-one relationship between redshift and frequency of dark photon/photon the decay dark photons never stimulate the decay. 
If we now relax this assumption and allow for a finite width of the decay profile, then the contribution of non-thermal background in the decay rate cannot be neglected. From Fig.~\ref{fig:temperature_comp} one finds that the non-thermal background is orders of magnitude higher than the thermal background. This immediately suggests that the self-stimulation effect could be very important and that significant departures from the thermal-background-only case can be expected.

In Appendix~\ref{sec:finite_width}, we derive the solution for the dark photon field including a finite-width profile. The solution is given by,
    \begin{equation}
    \label{eq:full_sol_taux}
    \Delta \tilde{N}_x=\frac{x^2}{c^2}\,[1+2n_{\rm bb}(x)]
    [\expf{\tau_x}-1]
     \end{equation}
and the optical depth is defined by Eq.~\eqref{eq:opt_tau_x}. The low optical depth limit directly recovers the previous solution, Eq.~\eqref{eq:sol_C22}, used above. However, in Fig.~\ref{fig:optical_depth} we see that the optical depth exceeds unity at all relevant dark photon frequencies. This means that the more general solution, Eq.~\eqref{eq:full_sol_taux} has to be used. This predicts a non-thermal dark photon background that is orders of magnitudes larger than what is require to explain the radio excess with the model parameters considered above. In particular, the associated radio background would be highly curved and far from a simple power law once self-stimulation is taking into account. 

For self-stimulation to be inefficient, the decay profile has to be narrow enough for the created dark photons to redshift out of the profile before the next dark matter particle is encountered. The average distance between dark matter particles is 
\begin{align}
L_{\rm X}\simeq N_{\rm X}^{-1/3}\approx 1.4\times 10^{-5}\,\left[\frac{m_X}{6\times 10^{-4}\,{\rm eV}}\right]^{1/3}\left[\frac{1+z}{1000}\right]^{-1}\,{\rm cm},
\end{align}
where $N_{\rm X}$ is the dark matter number density. The amount of redshifting one encounters while crossing this distance is 
\begin{align}
\delta \ln (1+z) \simeq \frac{L_{\rm X}}{L_H}\approx 1.7\times 10^{-29} 
\left[\frac{m_X}{6\times 10^{-4}\,{\rm eV}}\right]^{1/3}\left[\frac{1+z}{1000}\right]^{1/2}, 
\end{align}
where $L_H=c/H$ is the Hubble distance and we assumed matter-domination in the last step. Only if the decay width is smaller than this, would one be able to neglect self-stimulation. This is unlikely to be the case, as many processes are expected to broaden the effective decay profile, even if the intrinsic vacuum decay width may be extremely small. For example, the broadening caused by velocity dispersion in the standard cosmological context should be of the order of $\simeq 10^{-4}-10^{-3}$, which is well in excess of this. Therefore, it seems difficult to be able to ignore self-stimulation for this model in the range of parameters of interest, especially since the non-thermal dark photon background becomes so large in comparison to the thermal background. One may also have to take into account coherence effects if the mean separation between particles is comparable to the Compton wavelength. However, we do not pursue this complication here.

These findings suggest that a more careful consideration of the dark photon production process might be needed. In particular, the inverse (i.e., dark matter creation by dark photon absorption) process can likely no longer be neglected. In addition, elastic dark photon scattering terms may become relevant for thermalizing the dark photon field. We leave a study of these aspects to future work.

\section{Discussions and conclusion}
\label{sec:discussion}
In this paper, we have revisited the computation of C22 with detailed discussions about some of the aspects of the model. The goal was to clarify individual effects and to yield a refined understanding of the inherent assumption. It was suggested that by invoking a few ingredients (see introduction) one may be able to use an approximate power law [Eq.~\eqref{eq:temp_amplitude}] to describe the radio excess within the whole of frequency range 20~MHz--10~GHz. 
%
%
In this paper, we highlight that one expects significant departures from a pure power law. Indeed Eq.~\eqref{eq:temp_amplitude} is not a good description of the model, even approximately, over the frequency range of interest. Inclusion of radiation energy density and dark energy introduces inevitable curvature to the spectrum. Furthermore, the exact stimulated-decay lifetime and frequency-dependent probability of dark photon-to-photon conversion results in additional curvature at high frequency. 

Even if one finds significant departure from the power-law behaviour near $\simeq 1-10$ GHz, we note that the error bars in this frequency range are quite high and engulf the predicted departure. Available RSB data is indeed far more constraining around $\approx$ 20 MHz. We highlight that the radio spectrum is sensitive to contribution from the radiation energy density, at this frequency. We also find that the exact spectrum, actually, results in a better fit to the RSB data compared to a power-law spectrum in the absence of another source of radio photons such as the  extra-galactic radio background. Even with this improvement, we find that it is most likely that dark photons which today appear as $\simeq 10$ GHz photons are emitted at $z$ significantly less than 5. In that case, one cannot ignore the anisotropy induced by dark matter distribution which is correlated with large scale structure. With progress on measurement of radio background anisotropy, one can hope to constrain or even rule out the model in future \citep{Holder2014,OSHHL2022}. 

While we have concentrated on one particular model in this paper, more general conclusions can be drawn. Within a particle physics model such as dark matter decay/annihilation to radio photons or a cascade of particles injecting energy, or even astrophysical models, it is rather unlikely that the resulting photon spectrum is a pure power law over three decades of frequencies without any additional feature in the spectrum. As we argue here, using the expansion of the universe to fix the issue, cannot maintain a power law within a matter-dominated universe. One may think that it is possible to replace the matter era with the radiation era. But soft photons in the radiation era will be absorbed by the background electrons and will lead to appreciable CMB spectral distortions \citep{Chluba2015GreensII, Bolliet2020PI}, making this possibility difficult to realize. 

Our calculations also showcase the importance of going beyond the power law, which was originally used to obtain the best fit for RSB data \citep{ARCADE2011}. Deviation from a pure power law is already expected if in addition to the pure power law, one takes into account the contribution to radio background just from faint, extragalactic radio sources which have a spectral index $\gamma\simeq-2.7$. The combination of this contribution and the predicted signal from dark photons will lead to slight curvature, which does indeed lead to a better fit to RSB data. We note that in order to completely rule out the model that we consider here, one needs precise data at $\simeq~1-10$~GHz, which motivates a new ARCADE-like experiment. Combined with a CMB spectral distortion experiments such as {\it PIXIE} \citep{Kogut2011PIXIE, Kogut2016SPIE} or within the ESA Voyage 2050 program \citep{Chluba2021ExA}, this should allow us to test particle-physics inspired models for the observed radio excess.   

Finally, we demonstrated that even a tiny finite width of the dark matter decay profile might lead to a significant self-stimulation of the decay (see Sect.~\ref{sec:stim}). This is because the non-thermal dark photon background produced by the decay is actually not a small 'distortion' of the thermal dark photon background (see Fig.~\ref{fig:temperature_comp}). For exact $\delta$-function injection, the self-stimulation can be avoided, but exponential runaway is found for a finite width. Together with the expected anisotropy from the dark photon to photon conversion process, this suggests that additional work may be needed to obtain a viable model from the considered scenario.

\vspace{-3mm}
{\small
\section*{Acknowledgments}

We would like to thank Jack Singal and Andrea Caputo on their detailed comments on our draft.
This work was supported by the ERC Consolidator Grant {\it CMBSPEC} (No.~725456).
JC was furthermore supported by the Royal Society as a Royal Society University Research Fellow at the University of Manchester, UK (No.~URF/R/191023).
}

\vspace{-5mm}
\section{Data availability}
The data underlying in this article are available in this article and can further be made available on request.

{\small
\vspace{-3mm}
\bibliographystyle{mn2e}
\bibliography{Lit}
}

\appendix

\vspace{-3mm}
\section{Useful relations between dark matter mass and dark photon properties}
\label{sec:appendix}



The frequency of dark photons in rest frame for a given dark matter mass $m_X$ is given by,
\begin{equation}
    \nu_{A,0}=\frac{m_X c^2}{2h}=10\,{\rm GHz} \left[\frac{m_X}{8\times 10^{-5}\,{\rm eV}}\right].
\end{equation}
We require that we have injections of dark photons with $\nu_{A,0}$ such that it shows up as photons with frequency at least up to 10 GHz, today. It is useful to know the injection redshift of dark photons corresponding to these 10 GHz photons, which can be used as an anchor to calculate the injection redshift of any lower or higher frequency dark photons. The expression is then given by,
\begin{equation}
    1+z_{10 {\rm GHz}}=\frac{\nu_{A,0}}{10{\rm GHz}}.
\end{equation}
Knowing this, one can calculate, for example, the injection redshift of 20 MHz photons which is given by, $1+ 
z_{20 {\rm MHz}}=(1+z_{10 {\rm GHz}})\times 500$ ($10 {\rm GHz}/20 {\rm MHz}=500$). 

\vspace{-3mm}
\section{Dark photon spectrum}
\label{app:emission_solution}
The equation describing the evolution of the average number density of dark photons per unit frequency, in the expanding universe, is given by \citep{RybickiDell94, Chluba2008b},
\begin{equation}
    \frac{1}{c}\left[\partial_t N_{\nu}+2HN_{\nu}-H\nu\partial_{\nu}N_{\nu}\right]=\frac{\alpha\dot{\Gamma_X^* N_X}\,\delta(\nu-\nu_{0})}{4\pi},
    \label{eq:numdens_A}
\end{equation}
where $\nu$ is the frequency of dark photon at $z$, $\nu_{0}$ is rest frame frequency of the injected decay photon and $N_X$ is the number density of dark matter. Since we are only considering decays with lifetimes longer than the current age of the universe, $N_X a^3 \simeq {\rm const}$. Note that we have dropped the symbol "A" from the equation to avoid clutter and we will add it back in the final expression. In addition, we neglected any inverse process following the arguments of C22.
We next use the transformation, $x=a\nu$, which results in $N_{\nu}\id\nu=N_x\id x$ due to photon number conservation; Eq.~\eqref{eq:numdens_A}, then transforms to,
\begin{equation}
    \frac{1}{a^3}\frac{\partial(a^3N_x)}{\partial t}\Bigg|_x=\frac{c}{a}\frac{\alpha\dot{\Gamma^*_X N_X}\,\delta(x/a-\nu_{0})}{4\pi}.
    \label{eq:numdens_A1}
\end{equation}
Carrying out the transformations, $\tilde{N}_x=a^3N_x$ and from proper time to redshift, Eq.~\ref{eq:numdens_A1} becomes,
\begin{equation}
    \partial_z\tilde{N}_x=-\frac{{ c}}{H(z)}\frac{\alpha\Gamma_X^* \tilde{N}_X\,\delta(x/a-\nu_{0})}{4\pi},
    \label{eq:numdens_A2}
\end{equation}
where the symbol $\tilde{N}$ represents comoving quantities.
After performing the integral, we then have
\begin{equation}
    \Delta \tilde{N}_x(z)=\tilde{N}_x(z)-\tilde{N}_x(\infty)=\frac{{ c}\alpha\tilde{N}_X}{4\pi}\int_z^{\infty}\frac{\Gamma_X^*(z') }{H(z')}\delta\left(\frac{x}{a'}-\nu_0\right)\id z,
    \label{eq:numdens_A3}
\end{equation}
which then leads to
\begin{equation}
    \Delta \tilde{N}_x(z)=\frac{{c}\alpha}{4\pi x}\frac{\tilde{N}_X(z_*)\,\Gamma_X^*(z_*)}{H(z_*)}\Theta(z_*-z).
    \label{eq:APP_B6}
\end{equation}
This expression can be written in our notation as,
\begin{equation}
    \Delta \tilde{N}_{\nu_A}(z)=\frac{{c}\alpha}{4\pi \nu_A}\frac{\tilde{N}_X(z_*)\,\Gamma_X^*(z_*)}{H(z_*)}\Theta(z_*-z),
    \label{eq:numdens_A4}
\end{equation}
where $z_*$ is the redshift at which dark photons were injected from dark matter decay.

\section{Decay with finite width}
\label{sec:finite_width}
To include the effects of finite widths of the decay profile, we can start from Eq.~\eqref{eq:numdens_A2}, but have to replace the $\delta$-function with a general profile, $\phi(\nu)$. The profile is normalized such that $\int \phi(\nu)\id \nu=1$. This then yields the evolution equation
\begin{equation}
    \partial_z\tilde{N}_x=-\frac{{ c}}{H(z)}\frac{\alpha\Gamma_X \tilde{N}_X}{4\pi}\,\phi(x/a)\,\left(1+2\,\frac{c^2\tilde{N}_x}{2 x^2}\right).
    \label{eq:numdens_C1}
\end{equation}
The finite width of injection is parameterized by $\phi(\nu)$ which can be assumed to be a Gaussian as first approximation. However, we will show that our results are independent of assumed profile, as long as it is sufficiently narrow.

Using $\tilde{N}_x=\tilde{N}_{\rm bb}+\Delta \tilde{N}_x$ and $\partial_z \tilde{N}_{\rm bb}=0$, we then have
\begin{align}
    \partial_z\Delta \tilde{N}_x
    &=-\frac{{ c}}{H(z)}\frac{\alpha\Gamma_X \tilde{N}_X}{4\pi}\,\phi(x/a)\,\left\{[1+2n_{\rm bb}(x)]
    + \frac{c^2\Delta \tilde{N}_x}{x^2}\right\}
    \nonumber\\
    &=-\frac{\kappa(z)}{c^2}\,\phi(x/a)\,[1+2n_{\rm bb}(x)]-\frac{\kappa(z)}{x^2}\,\phi(x/a)\,\Delta \tilde{N}_x,
    \label{eq:numdens_C2}
\end{align}
where in the second line we introduced 
\begin{equation}
\kappa(z)=\frac{\alpha\Gamma_X \tilde{N}_X}{H(z)}\frac{{c}^3}{4\pi}.
\end{equation}
Defining the emission optical depth, $\tau_x(z)=\int_z^{\infty}\frac{\kappa(z')}{x^2}\phi(x/a')\id z'$, Eq.~\eqref{eq:numdens_C2} can be re-written as,
\begin{align}
    \partial_z\Delta \tilde{N}_x-\Delta \tilde{N}_x\partial_z\tau_x
    &\equiv\expf{\tau_x}\partial_z [\expf{-\tau_x}\Delta \tilde{N}_x]
    =\frac{x^2}{c^2}\,[1+2n_{\rm bb}(x)]\,\partial_z\tau_x.
    \nonumber
\end{align}
This solves to
\begin{align}
    \expf{-\tau_x(z)}\Delta \tilde{N}_x
    &=\frac{x^2}{c^2}\,[1+2n_{\rm bb}(x)]
    \int_\infty^z\expf{-\tau_x(z')}\partial_{z'}\tau_x(z') \id z'
    \nonumber\\
    &=\frac{x^2}{c^2}\,[1+2n_{\rm bb}(x)]\,
    [1-\expf{-\tau_x(z)}],
    \label{eq:numdens_C4}
\end{align}
which then implies the final solution at $z$ as
\begin{align}
    \Delta \tilde{N}_x
    &=\frac{x^2}{c^2}\,[1+2n_{\rm bb}(x)]\,
    [\expf{\tau_x}-1]\stackrel{\tau_x\ll 1}{\approx} \frac{x^2}{c^2}\,[1+2n_{\rm bb}(x)]\,\tau_x,
    \label{eq:numdens_C5}
\end{align}
where we gave the small optical depth limit in the second step.

For a narrow decay profile, most of the contribution to $\tau_x$ comes from $x/a_*=\nu_0$. Since the redshift-dependent coefficient is very smooth function of $z$, it can be taken out of integral. Therefore, the expression for $\tau_x$ simplifies to,
\begin{equation}
    \tau_x(z)\approx \frac{\kappa(z_*)}{x^2}
\int_{z}^{\infty}\phi(x/a')\id z'
\end{equation}
Since, $\nu=x(1+z)$ and $\phi(x/a)$ is normalized, the expression for $\tau_x$ finally becomes,
\begin{equation}
\label{eq:opt_tau_x}
    \tau_x(z)\approx \frac{\kappa(z_*)}{x^3}\,\Theta(z_*-z)
\end{equation}
with $x=\nu_0/(1+z_*)$. This result is independent of the exact profile, as promised. Inserting this into the low-optical depth limit, we find
\begin{align}
    \Delta \tilde{N}_x\approx 
    \frac{\kappa(z_*)}{c^2 x}\,[1+2n_{\rm bb}(x)]\,\Theta(z_*-z)
    =
    \frac{{c}\alpha}{4\pi x}\frac{\Gamma^*_X(z_*) \tilde{N}_X(z_*)}{H(z_*)}\,\Theta(z_*-z),
    \nonumber
\end{align}
which is the previous solution, Eq.~\eqref{eq:APP_B6}, which led to Eq.~\eqref{eq:sol_C22} for a $\delta$-function profile. However, it turns out that for the relevant parameters $\tau_x\gg 1$ at the relevant frequencies. This means that we instead see an exponential runaway, as discussed in Sect.~\ref{sec:stim}.

\end{document}